\documentclass[11pt]{article}
\pdfoutput=1

\usepackage{multirow}
\usepackage{amsmath, amsfonts, amssymb}
\usepackage{comment}
\usepackage{graphicx}
\usepackage{psfrag}
\usepackage{amsthm}
\usepackage[usenames,dvipsnames,svgnames,table]{xcolor}
\usepackage{enumerate}
\usepackage{diagbox}

\usepackage{soul}
 \usepackage{slashed}
\usepackage{subfig}

\usepackage{mathrsfs}  
 \usepackage{a4wide}
  \usepackage{tikz}
  \usepackage{tikz-cd}

  \usepackage{color}
  \definecolor{dark-gray}{gray}{0.20}
  \definecolor{gray}{gray}{0.30}
  \definecolor{light-gray}{gray}{0.80}
  \definecolor{dark-red}{rgb}{0.7,0,0}
  \definecolor{dark-green}{rgb}{0.1,0.4,0}
  \definecolor{dark-blue}{rgb}{0.3,0.3,0.7}
  \definecolor{light-blue}{rgb}{0.8,0.8,1}
   \definecolor{swamp}{RGB}{240, 199, 197}
     \definecolor{landscape}{RGB}{180, 250, 199}
          \definecolor{undecided}{RGB}{252, 252, 197}
  
  \usepackage{pifont}
\newcommand{\cmark}{\ding{51}}%
\newcommand{\xmark}{\ding{55}}%

\newcommand{\mcmark}{\text{\ding{51}}}%
\newcommand{\mxmark}{\text{\ding{55}}}%

\usepackage{setspace}
\newcommand{\be}{\begin{equation}}
\newcommand{\ee}{\end{equation}}
\newcommand{\eq}[1]{(\ref{#1})}

  \usepackage{jheppub}
\hypersetup{
	colorlinks=true,
	linkcolor=dark-blue,
	citecolor=dark-red,
	urlcolor=dark-green,
	linktoc=page
}

\def\be{\begin{equation}}
\def\ee{\end{equation}}
\def\bea{\begin{eqnarray}}
\def\eea{\end{eqnarray}}

\def\simleq{\; \raise0.3ex\hbox{$<$\kern-0.75em
      \raise-1.1ex\hbox{$\sim$}}\; }
   \def\simgeq{\; \raise0.3ex\hbox{$>$\kern-0.75em
      \raise-1.1ex\hbox{$\sim$}}\; }

\numberwithin{equation}{section}

\title{\centering Cobordism Conjecture, Anomalies, and the String Lamppost Principle}

\author{Miguel Montero}
\author{and Cumrun Vafa}

\affiliation{Jefferson Physical Laboratory, Harvard University,\\
Cambridge, MA 02138, USA}
\emailAdd{mmontero@g.harvard.edu, vafa@g.harvard.edu}

\abstract{We consider consequences of triviality of cobordism classes and anomaly cancellation in supergravity theories in $d>6$.
We argue that this leads to the existence of certain defects which we call ``I-folds'' (a generalization of orientifolds).  The requirement that compactifications to lower dimensions involving these defects be anomaly free leads to conditions on the higher dimensional theory.  We show that in theories with 16 supercharges in $d>6$ this leads to restrictions on the rank of the allowed gauge groups and thus provides an explanation for the observed restrictions in known string theory constructions.  In particular, in eight and nine dimensions the only solutions to our constraints are precisely the ones realized in string theory compactifications. We also use these techniques to place constraints on the global structure of the gauge group in eight and nine dimensions.
}

\begin{document}

\makeatletter
\let\old@fpheader\@fpheader

\makeatother

\maketitle

\section{Introduction}\label{sec:intro}

One of the key questions in high-energy physics is what constitutes a consistent quantum theory of gravity.  Understanding this systematically is the basic aim of the Swampland program \cite{Vafa:2005ui} (see also the reviews \cite{Brennan:2017rbf,Palti:2019pca}).  String theory constructions provide a laboratory of features that a consistent quantum gravitational theory should enjoy.  To narrow down the search for Swampland principles it is natural to restrict to the most symmetric cases in string theory, and that would involve constructions with high supersymmetry in higher dimensions admitting Minkowski backgrounds.  The low energy content of theories with the highest amount of supersymmetry (32 supercharges) is uniquely fixed by supersymmetry.  The next case of interest are theories with 16 supercharges, which is the only other possibility in $d>6$.  

Progress in this direction has been made recently.  In particular it has been shown that the only possibility in $d=10$ are the ones with gauge groups $E_8\times E_8$ and $SO(32)$ \cite{Adams:2010zy,Kim:2019vuc}. For $d<10$ it has been shown that the rank of the gauge group is bounded by $r_G\leq 26-d$ \cite{Kim:2019ths}, which is consistent with all known string constructions.  This in particular establishes that only a finite number of matter contents are possible, at least for theories with 16 supercharges. However there is more structure in what is observed in string theory constructions.  In particular, not all ranks in the allowed range seem to appear.  For example in $d=9$, the only observed ranks are $1,9,17$.  It is important to settle whether there are
other theories with rank less than or equal to 17  in $d=9$ in addition to the known string compactifications.  If there are and string theory constructions cannot yield them, it shows the incompleteness of the string landscape.  On the other hand if these rank restrictions are forced by consistency arguments of quantum gravity alone, it lends further support for the String Lamppost Principle (SLP):  That the string landscape includes all consistent quantum gravitational backgrounds.  One aim of this paper is to show that the rank restriction at least in higher dimensions is a necessary consequence of other well established swampland principles, such as the absence of global symmetries.  

One proposed Swampland principle we use is the Cobordism Conjecture \cite{McNamara:2019rup}.  This conjecture, which can be viewed as a generalization of the lack of global symmetries and completeness of spectrum in quantum theories of gravity (in $d>3$), states that in a consistent theory of gravity all cobordism classes are trivial.  In particular if there appears to be a non-trivial cobordism class at low energies, there must exist an object within the theory which trivializes that class.  We use this conjecture to argue that in any consistent quantum theory of gravity there must exist defect singularities we call ``I-folds'' (inversion-folds), which are a generalization of orientifold planes.  Using this we construct compactifications of a putative consistent quantum gravity theory with 16 supercharges to lower dimensions with 8 supercharges.  Consistency of the lower dimensional theory, and in particular anomaly cancellation, turns out to require that the higher dimensional theory satisfy some rank conditions.  In particular we show that in $d=9$ the rank should be 1 mod 8, in $d=8$ it should be 2 mod 8, and $d=7$ should have odd rank.  All these are consistent with rank restrictions arising in string theory constructions and for $d=8,9$ it yields exactly the observed ranks in string compactifications.

The organization of this paper is as follows.  In Section \ref{sec:CP} we review the rank restrictions in $d>6$ theories from known string constructions.  In Section \ref{sec:main} we review ${\sf P}$ and ${\sf CP}$ symmetries of supergravity theories and explain how the cobordism conjecture leads to the existence of I-folds.  We then use these ingredients in compactification to lower dimensions and use anomaly cancellation in the lower dimensional theory to obtain rank restrictions for the higher dimensional theory.  In Section \ref{sec:P} we discuss constraints on the global structure of the gauge groups that can appear at points of enhanced symmetry.  In Section \ref{sec:lattices} we discuss aspects of charge lattices that can or cannot arise in theories with 16 supercharges.  We conclude in Section \ref{sec:conclus} with  some open questions.  Some of the technical details of the arguments are presented in the appendices.

\section{Theories with 16 supersymmetries and string constructions for \texorpdfstring{$d>6$}{d>6}}\label{sec:CP}

In this section  we will start with a brief review of the data required to specify a theory with $16$ supercharges, and their realization in string theory in dimensions $d>6$.  The case with 10 dimensions is already well known so we focus only on dimensions $d=7,8,9$.

\subsection{\texorpdfstring{$d=9$}{d=9}}
 Consider a nine-dimensional  $\mathcal{N}=1$ (16 real supercharges) supergravity theory. We work in Lorentzian signature. There are only two multiplets in nine dimensions \cite{Aharony:2007du}:
\begin{itemize}
\item Gravity multiplet: Contains the graviton, a 2-form, the gravitino, dilatino (both Majorana), a $U(1)$ graviphoton and a real  scalar.
\item Vector multiplet: Contains a vector, a Majorana fermion, and a scalar in the adjoint representation of the group. 
\end{itemize}
The scalars in the vector multiplets parametrize the gauge kinetic function of the vectors. The gravity multiplet scalar contains a ``dilaton'' that doubles down as the gauge kinetic function of the vector in the gravity multiplet. 

At the supergravity level, the action enjoys an $O(1,r;\mathbb{R})$ symmetry \cite{Ortin:2015hya}, (where $r$ is the rank of the matter gauge group) which in general is broken to a discrete subgroup via quantum effects. We will have more to say about this in Section \ref{sec:P}. On top of this, the supersymmetry algebra enjoys a discrete $\mathbb{Z}_2$ ``R-symmetry'' that multiplies the supercharges by $-1$. This is just $(-1)^F$, global fermion number. 

The scalars cannot get a potential due to supersymmetry, and locally they parametrize a Narain moduli space. On a generic point in moduli space, the gauge group (including the graviphoton) is simply
\begin{equation}U(1)^{r+1}.\end{equation}
 At special points in moduli space, additional vectors become massless and one can have enhancement to a non-abelian gauge group \cite{polchinski1998string}. In fact, only the Lie algebra is relevant at the supergravity level, since the scalars in the vector multiplets are charged under the adjoint representation of the gauge group. Since adjoint breaking always preserves the rank of the gauge group, this cannot change as we move continuously in the moduli space. 
 
 Thus, the moduli space of all consistent $\mathcal{N}=1$ theories of quantum gravity in nine dimensions splits into a number of disconnected components, labelled by the rank and possibly additional discrete labels which are invisible from the supergravity point of view\footnote{This happens for instance in the known rank 1 theories, for which stringy embeddings provide two disconnected components in the moduli space \cite{Aharony:2007du,Kim:2019ths}.}.

To fully specify the low energy limit of a consistent supergravity in nine dimensions, we only need two pieces of data:\begin{itemize}
\item The rank $r$ of the gauge group, and
\item a list of the possible Lie algebras that can arise at points of enhanced symmetry. 
\end{itemize}
The Swampland program for 16 supercharges in nine dimensions is therefore just about figuring out precisely which combinations of the above have consistent UV completions. If one were able to show that everything that can be obtained from known string constructions matches what is allowed from Swampland constraints, it would be evidence for the String Lamppost Principle-- that every consistent 9d $\mathcal{N}=1$  supergravity is realized in string theory.  

So what is known so far in the 9d case based on general consistency requirements of quantum gravity? We have some constraints for both kinds of data above:
\begin{itemize}
\item The rank $r$ has to be odd, otherwise, there is a global gravitational anomaly \cite{AlvarezGaume:1983ig}. Thus, in particular, pure supergravity is in the Swampland.
\item In \cite{Kim:2019ths}, an upper bound for the rank of the gauge group was derived from a strong form of Swampland distance conjecture. The upper bound in 9d is
\begin{equation} r\leq 17.\end{equation}
\item In \cite{Garcia-Etxebarria:2017crf}, it was proven that the $\mathfrak{f}_4$,  and  $\mathfrak{b}_n$ Lie algebras have a global gauge anomaly and therefore are inconsistent in eight dimensions, and so also in nine  \footnote{$\mathfrak{c}_n$ has an anomaly of a different kind in eight dimensions that does not render the theory inconsistent. }.
\end{itemize}
These constraints are not sufficient to yield what we can actually get from string theory.   See \cite{Aharony:2007du,Kim:2019ths} for a survey of string theories in 9, 8 and 7 dimensions.  Let us focus on a generic point in the Coulomb branch,  away from any enhanced symmetry locus. The constraints listed above allow for any rank up to 17, while in string theory we only know how to construct
\begin{equation}r=1,9,17.\label{ranks9d}\end{equation}
 There is a single theory at rank 17, which can be obtained as a compactification of heterotic to 9d; one rank 9 theory, the CHL string \cite{Chaudhuri:1995fk,Chaudhuri:1995bf}; and two disconnected, inequivalent rank 1 theories, coming from M-theory on the Klein Bottle and IIB on the Dabholkar-Park background \cite{Dabholkar:1996pc}. Notice the fact that the ``gaps'' in the rank of existing 9d theories are multiples of 8. Explanation of ``modulo 8'' periodicity of the rank of known string theories is one goal of this paper.

The situation regarding the possible Lie algebras that can arise in 9d is similar: There is also a mismatch between the known Swampland constraints and what we can get from string theory. For simplicity, let us restrict the question to what semi-simple Lie algebras can arise; then, $\mathfrak{g}_2$ and $\mathfrak{c}_n$ are conspicuously absent from the list of known stringy realizations, but are not forbidden by any Swampland constraint either. We will comment on this in Section \ref{sec:P}.

\subsection{\texorpdfstring{$d=8$}{d=8}}
Eight-dimensional theories with sixteen supercharges are in many ways very similar to their nine-dimensional counterparts. The  reason is that the supercharges transform in a sixteen-dimensional Majorana representation. Again, there are just gravity and vector multiplets. The main difference is that the scalar in the vectors is complex and that there is a dilatino and two graviphotons rather than just one in the gravity multiplet. Although local anomalies are possible in 8d, they all cancel automatically in supersymmetric theories. The same comments about the moduli space apply, and again the only data required to specify the low energy limit of a theory is a number $r$ (the rank of the gauge group) and a list of the possible non-abelian enhancements.  See \cite{Taylor:2011wt,Kim:2019ths} for reviews.

The known Swampland constraints on the rank are:\begin{itemize}
\item There is still a global anomaly \cite{AlvarezGaume:1983ig,Witten:1985xe} which now restricts the rank to be even. 
\item Again, there is an upper bound on the rank from the considerations in \cite{Kim:2019ths},
\begin{equation} r\leq 18.\end{equation}
\end{itemize}
The constraints on the algebra are the same as in 9d. The only known constructions in string theory ar obtained from dimensional reduction of the nine-dimensional ones. In particular, we have
\begin{equation}r\equiv 2,10,18.\end{equation}
We see there is again a modulo 8 periodicity, that we will explain later on. Unlike in nine dimensions, symplectic groups with algebra $\mathfrak{c}_n$ are now present, although only up to $Sp(10)$ \cite{Chaudhuri:1995bf}; we lack a construction for $Sp(n)$ for $n> 10$.  The only other exception is $\mathfrak{g}_2$, for which there is no known string construction either. Other than these, all semi-simple Lie algebras with rank in allowed range and without global anomalies arise somewhere in moduli space.

\subsection{\texorpdfstring{$d=7$}{d=7}}
Seven-dimensional theories are richer than their higher-dimensional counterparts. They are again specified by the same set of data, but this is significantly less constrained. There is an upper bound on the rank from \cite{Kim:2019ths}, which is now
\begin{equation} r\leq 19,\end{equation}
but no constraint on the parity of the rank coming from global gravitational anomalies. All exceptional semi-simple Lie algebras and all infinite families arise now at some point of the moduli space from stringy constructions \cite{deBoer:2001wca, Aharony:2007du}, including $\mathfrak{g}_2$. The values of the rank that we know how to get from string compactifications are
\begin{equation}r=\, 3,5,7,11\, \text{and}\, 19.\end{equation}
The rank is odd, but there seems to be more structure.  There is no simple periodicity in the rank, but note the tantalizing facts that the increases in rank are always a power of two, and that the ranks are all prime numbers. We cannot hope to explain the full structure of the observed ranks with the simple techniques used here, but we will be able to explain why there are no theories of even rank.  It would also be interesting to see if there are additional string constructions which give all the allowed odd ranks.

The constraints  on 7,8, and 9d theories that we discussed here are succinctly summarized in Table \ref{tabla1}.

\begin{table}[!hbt]
\begin{center}
\begin{tabular}{c|cccccccccccccccccccc}
$d$&\multicolumn{19}{c}{$r$}\\\hline
$9$&\cellcolor{swamp}0&\cellcolor{landscape}1&\cellcolor{swamp}2&\cellcolor{undecided}3&\cellcolor{swamp}4&\cellcolor{undecided}5&\cellcolor{swamp}6&\cellcolor{undecided}7&\cellcolor{swamp}8&\cellcolor{landscape}9&\cellcolor{swamp}10&\cellcolor{undecided}11&\cellcolor{swamp}12&\cellcolor{undecided}13&\cellcolor{swamp}14&\cellcolor{undecided}15&\cellcolor{swamp}16&\cellcolor{landscape}17&\cellcolor{swamp}&\cellcolor{swamp}\\\hline
$8$&\cellcolor{undecided}0&\cellcolor{swamp}1&\cellcolor{landscape}2&\cellcolor{swamp}3&\cellcolor{undecided}4&\cellcolor{swamp}5&\cellcolor{undecided}6&\cellcolor{swamp}7&\cellcolor{undecided}8&\cellcolor{swamp}9&\cellcolor{landscape}10&\cellcolor{swamp}11&\cellcolor{undecided}12&\cellcolor{swamp}13&\cellcolor{undecided}14&\cellcolor{swamp}15&\cellcolor{undecided}16&\cellcolor{swamp}17&\cellcolor{landscape}18&\cellcolor{swamp}\\\hline
$7$&\cellcolor{undecided}0&\cellcolor{undecided}1&\cellcolor{undecided}2&\cellcolor{landscape}3&\cellcolor{undecided}4&\cellcolor{landscape}5&\cellcolor{undecided}6&\cellcolor{landscape}7&\cellcolor{undecided}8&\cellcolor{undecided}9&\cellcolor{undecided}10&\cellcolor{landscape}11&\cellcolor{undecided}12&\cellcolor{undecided}13&\cellcolor{undecided}14&\cellcolor{undecided}15&\cellcolor{undecided}16&\cellcolor{undecided}17&\cellcolor{undecided}18&\cellcolor{landscape}19\\
 \end{tabular}
\end{center}

\caption{State of the art on constraints on the rank of theories with sixteen supercharges in $d>6$ prior to this paper. In red, theories that are known to be inconsistent. In green, theories which have a known stringy embedding. In orange, theories which are not ruled out, but which have no known string embedding. Ranks above $9+d$ are excluded by \cite{Kim:2019ths} and so are not displayed.}
\label{tabla1}
\end{table}

\section{Rank restrictions and cobordism triviality}\label{sec:main}
We now come to the main point of this note, which is that rank restrictions are actually a consequence of the cobordism conjecture together with anomaly cancellation. The key assumption will be that theories with 16 supercharges admit non-orientable manifolds as consistent backgrounds, and in particular $\mathbb{RP}^2$ is a consistent background. This works because these supergravity theories all admit an action of parity as an internal symmetry. We first discuss this symmetry in Subsection \ref{sub:par}. Then we explain the main argument in the nine-dimensional case in Subsection \ref{sec:Ifolds}. Subsequently, in Subsections \ref{sec:9dmod8}, \ref{sec:8dmod8} and \ref{sec:7dmod2}, we apply it to 9d, 8d, and 7d theories, respectively. 

\subsection{Parity symmetry}\label{sub:par}
We study parity symmetries of $d>6$ supergravity theories with 16 supercharges. These theories admit parity symmetries, and to fully specify the action of a parity symmetry in the low energy limit of supergravity, we just need to specify how it acts on the light fields. As explained e.g. in \cite{Witten:2015aba},  on general grounds parity transformations act by Clifford multiplication on the fermions, and there are actually two different double covers of $O(d,1)$ on which the fermions can live, distinguished by whether 
 \begin{equation}{\sf P}^2=+1,\quad \text{($\text{Pin}^+$ structure)}\quad\text{or}\quad {\sf P}^2=(-1)^F,\quad \text{($\text{Pin}^-$ structure)}.\end{equation}
In general, the choice of whether to take $\text{Pin}^+$ or $\text{Pin}^-$ for a given theory is a physical one -- it affects on which manifolds the theory can be placed consistently, which internal symmetries are allowed, etc. 

\subsubsection{Parity symmetry in \texorpdfstring{$d=9$}{d=9}}
  In the particular case of a 9d $\mathcal{N}=1$ theory, the fact that the supercharge (as well as gravitino and gaugini) transform in the 16-dimensional Majorana representation is only compatible with a parity symmetry that squares to $-1$ on fermions. This is explained in detail in Appendix \ref{app:ferms}. In other words,
\begin{center}
{$\mathbf{\mathcal{N}=1}$ theories in nine dimensions are only compatible with a $\text{Pin}^-$ structure.}
\end{center}
Once we know how the supercharge transforms, we just need the transformation properties of the primary fields in each supermultiplet \cite{Taylor:2011wt}, to determine how every field in the multiplet transforms. Massless multiplets in 9d are constructed in terms of the 7d  little group. Raising operators that can be constructed out of the supercharge transform in the $\mathbf{1}_++ \mathbf{8}+\mathbf{7}_-$, where the $\pm$ indicates whether the transformation acts as ordinary parity or the field picks an additional minus sign. From this we get that \begin{itemize}
\item The gravity multiplet, built out of the massless $\mathbf{7}_-$, has a dilaton, $B$-field and metric transforming as ordinary tensors, while the one-form transforms as a pseudovector. Constructing the  multiplet out of $\mathbf{7}_+$ would lead to the dilaton flipping sign under parity, which would not be a symmetry of the lagrangian. 
\item One can construct two kinds of vector multiplets, acting with the supersymmetry generators on the $\mathbf{1}_\pm$. The bosonic content of one is a pseudoscalar and an ordinary vector; for the other, it is an ordinary scalar and a pseudovector.\end{itemize}

With the above assignments and for any choice for the vectors, the parity transformation is a symmetry of the 9d supergravity Lagrangian. However, if there is any vector multiplet whose scalar $\psi$ transforms nontrivially, the corresponding parity symmetry will be spontaneously broken everywhere on the moduli space except at the particular point $\psi=0$. We are led to the conclusion that there is a unique choice of parity transformation that is a symmetry everywhere in moduli space, where all the scalars transform trivially. We will call this symmetry ${\sf CP}$, since all vectors transform as pseudovectors, picking an action of charge conjugation; a massive state with charge $q$ would get mapped to a state of charge $-q$.

 The basic assumption underlying our argument is that this ${\sf CP}$ symmetry, which is a symmetry of the low energy Lagrangian, is  actually an exact symmetry of the $\mathcal{N}=1$ theory (and is not broken by massive excitations in the theory).  An equivalent statement would be that any consistent $\mathcal{N}=1$ theory makes sense on non-orientable manifolds, on a generic point of its moduli space. And yet another equivalent form of the statement is that any consistent S-matrix with $\mathcal{N}=1$ supersymmetry in nine dimensions is actually ${\sf CP}$-invariant. 
 
It may be that the existence of an exact parity symmetry is somehow a consequence of the supersymmetry algebra, though we do not have a proof. In this paper we will assume that ${\sf CP}$ symmetry is not broken due to massive excitations.  There may be a deeper reason for the existence of this symmetry, and we will speculate about this in Section \ref{sec:lattices};  for now, we will content ourselves with showing that it is true in all stringy examples. We will first construct a purely geometric, ${\sf P}$ symmetry, under which some of the scalars are odd; and then compose with an internal gauge transformation, which is always an exact symmetry of the theory, to form the  ${\sf CP}$ symmetry we are looking for. 

One can construct the ${\sf P}$ symmetry of $\text{Pin}^-$ type in any $d$-dimensional theory that arises as a circle compactification of a $(d+1)$-dimensional theory invariant under diffeomorphisms. Namely, consider the $d$-dimensional theory on a manifold $X_d$, which comes with an orientation-reversing involution $\tau$. Then, we can extend this to an involution of $X_d\times S^1$ that acts as
\begin{equation}(x,\theta)\,\rightarrow\, (\tau(x),-\theta)\label{s11}\end{equation}
This combined diffeomorphism is orientation-preserving, and so a symmetry of the original theory. When restricted to massless fields (the KK zero modes) on $X_d\times S^1$, however, it acts as a parity symmetry. On fermions, the symmetry in \eq{s11} acts as
\begin{equation} \psi(x,\theta)\,\rightarrow\, \hat{\tau}\,\hat{r}\,\psi(\tau(x),-\theta),\end{equation}
as long as there are no Wilson lines around $S^1$, where $\hat{\tau}$ is the lift of the $\tau$ action to the spin bundle and $\hat{r}$ is the lift of the circle reflection. If the fermions are real, it satisfies $\hat{r}^2=1$.  Squaring and anticommuting, one then gets that in order to have nonvanishing spinors we must impose 
\begin{equation} \hat{\tau}^2=-1,\end{equation}
so the symmetry is of $\text{Pin}^-$ type. If $\tau$ is fixed-point free, one could quotient $X_d\times S^1$ by the action \eq{s11}, producing a compactification\footnote{We are neglecting effects such as e.g. tadpole cancellation, which could force the introduction of fluxes or branes.} of the $d$-dimensional theory on the non-orientable $\text{Pin}^-$ manifold $X_d/\mathbb{Z}_2$. This shows that parity is gauged in this theory.

These constructions apply readily to the branches of known 9d string theory constructions, since they can all be described as backgrounds of the form $X_9\times S^1$ in some corner of the moduli space:\begin{itemize}
\item The rank 17 theories arise directly as circle compactifications of type I or the two heterotic 10d theories. 
\item The CHL can be described as a compactification of the $E_8\times E_8$  with a $\mathbb{Z}_2$ involution that exchanges the two $E_8$ factors; this does not affect the argument above, as the action \eq{s11} preserves this involution. 
\item Both branches of the moduli space of rank 2 theories have weakly coupled points where the theory can be described as asymmetric orbifolds of type II theories \cite{Aharony:2007du}.  An asymmetric orbifold is basically a circle compactification with a holonomy for a discrete $\mathbb{Z}_2$ symmetry (left-moving worldsheet fermion number), and so it is again preserved by \eq{s11}. 
\end{itemize}

Since the gauge bosons transform as ordinary vectors, it makes sense to call inversion symmetries of this kind ``${\sf P}$ actions''\footnote{Although notice that for the rank 2 theories, ${\sf P}={\sf CP}$.}. This notation also correctly suggests that to obtain the ${\sf CP}$ action we were looking for, we just need to include an additional ``charge conjugation'' transformation in the quotient in \eq{s11}. In particular, we could specify that the $\mathbb{Z}_2$ includes an action under any internal symmetry, such as a gauge transformation of the $\text{Spin}(32)/\mathbb{Z}_2$ or $E_8\times E_8$ gauge groups that arise at the fixed loci. In particular, there is a $\mathbb{Z}_2$ gauge transformation in $E_8$ that just flips the sign of all the Cartans and maps each weight vector to minus itself; by including this action in \eq{s11}, we turn ${\sf P}$ into our desired ${\sf CP}$ action, which acts in the same way over every vector, and that leaves a generic point in the moduli space invariant. 

\subsubsection{Parity symmetry in \texorpdfstring{$d=8$}{d=8}}\label{sec:par8d0}
Much of the discussion from the 9d case carries over to the 8d case. In particular, the supergravity still admits a $\text{Pin}^-$ parity symmetry (and only of this kind, as explained in Appendix \ref{app:ferms}). However, there are some important differences which we now discuss in detail. 

Fermion representations in eight dimensions are similar to four in that fermions can be Majorana or Weyl, but not both. The 8d gauginos and supercharge live in a 16-dimensional Majorana representation. It is really ``the same'' representation as the 9d Majorana, where one just chooses to forget about one coordinate. Unlike in 9d, where the only internal symmetry of the Majorana representation is multiplication by $-1$,  the 8-dimensional supergravity has an internal symmetry ${\sf \tilde{C}}$ (not to be confused with the charge conjugation ${\sf C}$ we introduced before) given by 
\begin{equation} \psi(x)\rightarrow \Gamma^9\,\psi(x),\label{errr}\end{equation}
where $\Gamma^9$ is the 9d real chirality matrix, which squares to $-1$. If we obtain an 8d theory from a circle compactification of a 9d one with a parity, \eq{errr} just tells us that a parity transformation in the 9th direction is an internal symmetry from the 8d point of view. In fact, as described in Section \ref{sec:CP}, all known 8d theories arise from 9d ones in this way. 

Just as we did with parity in nine dimensions, \eq{errr} is a symmetry of the supergravity fields which we will assume is not broken by massive states, and so it is a valid internal symmetry of any consistent 8d theory with 16 supercharges. This is true in all string examples, and we will provide some justification for it in Section \ref{sec:lattices}. Supersymmetry means that ${\sf \tilde{C}}$ must act by multiplication by $\pm1$ on the scalars of the vector multiplets. We will assume it is a symmetry on a generic point of the 8d moduli space, which means it will act as multiplication by $+1$.

The fact that there is an internal symmetry like ${\sf \tilde{C}}$ means that the action of a parity symmetry on the supercharges is no longer unique. A reflection along the $i$-th coordinate could act by multiplication by $\Gamma^i$, as in 9d, or it could be composed with ${\sf \tilde{C}}$, acting like $\Gamma^9\Gamma^i$. Thus we get two kinds of parity symmetry, which we will call ${\sf P}$ (the one that acts by multiplication by $\Gamma^i$, like its 9d counterpart) and ${\sf \tilde{C} P}$. The complex scalars in the vector multiplet transform as charge conjugation under either a ${\sf P}$ or a ${\sf \tilde{C} P}$ transformation. The easiest way to see this is to notice that the 8d multiplets are the 9d multiplets in disguise, and so the transformation properties can be readily obtained by dimensional reduction.

Since the complex scalars charge-conjugate, parity is only unbroken on a real submanifold.  This is unlike in 9d; now, for \emph{any} parity action (for instance, ${\sf P}$), we will be able to find a real submanifold where it remains unbroken.  On such a real locus the 8d theory has two parity symmetries, ${\sf P}$ and ${\sf \tilde{C} P}$.  Both ${\sf P}$ and ${\sf \tilde{C} P}$ square to $(-1)^F$ (they provide $\text{Pin}^-$ structures) and act in the same way on bosons on the real submanifold, but they act differently on fermions and, more importantly, they commute. This fact will play an essential role in Section \ref{sec:8dmod8}. 
The condition that there be a sublocus in moduli space of supersymmetric theories with enough supercharge where parity is unbroken was also noted in the context of ${\cal N}=2$ theories in 4 dimensions in \cite{Cecotti:2018ufg}. It is conceivable that this may be a general Swampland criterion. 
 
As an example, consider heterotic at the point of $\text{Spin}(32)/\mathbb{Z}_2$ symmetry. Compactifying on a circle, there is a ${\sf P}$ symmetry, acting as a reflection on the internal $S^1$, which no longer exists once Wilson lines are turned on. By contrast, compactifying on $T^2$ one can turn on a Wilson line on the cycle transverse to the one that is being reflected, and ${\sf P}$ is still a good symmetry\footnote{In fact, we have \emph{two} ${\sf P}$ symmetries at the unbroken point, but one or the other is broken as soon as we turn on Wilson lines.}.

\subsubsection{Parity symmetry in \texorpdfstring{$d=7$}{d=7}}\label{sec:par7d}
In seven dimensions, fermions must be Dirac, and there is a natural $U(1)$ R-symmetry (which will often be broken in stringy embeddings). Now both $\text{Pin}^+$ and $\text{Pin}^-$ structures are possible, although only $\text{Pin}^+$ can be unbroken for general choices of the Wilson lines. In e.g. heterotic on $T^3$, the $\text{Pin}^+$ structure is obtained by reversing the three internal coordinates as well as charge-conjugating the scalars coming from the 10d vectors; the $\text{Pin}^-$ structure is obtained by reflecting just one internal coordinate, and it will not be a symmetry at generic points in moduli space.

In Section \ref{sec:7dmod2}, we will derive a mod 2 condition on the rank of 7d theories, for which either $\text{Pin}^+$ or $\text{Pin}^-$ work. On a special locus of the moduli space where both symmetries are available, the symmetry group is enhanced to $\text{Pin}^{\mathbb{Z}_4}$; in this case, we can recover the mod 2 constraint in a slightly different way. 

\subsection{Cobordism conjecture and I-folds}\label{sec:Ifolds}

We are now in a position to apply general argument explained in the introduction to the particular case of 16 supercharge theories in dimensions $d=9, 8$ and $7$. Since they make sense on $\text{Pin}^-$ manifolds, the relevant bordism is $\text{Pin}^-$  (see \cite{Garcia-Etxebarria:2018ajm,McNamara:2019rup} for tables of different bordism groups). In particular,
\begin{equation}\Omega_2^{\text{Pin}^-}=\mathbb{Z}_8,\quad\text{generated by}\quad\mathbb{RP}^2.\label{bord0}\end{equation}
This means, due to cobordism conjecture, that any 9d $\mathcal{N}=1$ supergravity theory carries a $\mathbb{Z}_8$ global charge, which must be broken by some UV effect. That is, there must be some configuration $Y_3$, which cannot possibly be a smooth $\text{Pin}^-$ manifold, such that 
\begin{equation}\mathbb{RP}^2=\partial Y_3\quad\text{in}\quad \Omega_3^{\text{QG}}=0.\label{bord3}\end{equation}
In known $\mathcal{N}=1$ theories obtainable from string theory, $Y_3$ is geometrically an orbifold
\begin{equation}\mathbb{R}^3/\mathbb{Z}_2,\label{geom-orb}\end{equation}
typically accompanied by an additional internal action. The corresponding object has a $\mathbb{Z}_8$ gauge charge inherited from the fact that it is killing a $\mathbb{Z}_8$ cobordism class.  The fact that the charge can be measured from far away is a sign that it is gauged and should cancel in a compact space.   Once one has an object like \eq{geom-orb}, it is possible to glue together 8 copies to produce a compactification
\begin{equation}T^3/\mathbb{Z}_2.\end{equation}
This will be a 6d model with $\mathcal{N}=(1,0)$ or $\mathcal{N}=0$ supersymmetries. In both cases, anomalies will have to cancel.  

The existence of the objects of the form \eq{geom-orb} predicted by the cobordism conjecture is, of course, automatic in known string compactifications. As we will see momentarily, the description of the relevant objects sometimes involves orientation reversal (as in type I) and sometimes it does not (as in the heterotic frame). Since in any case they always involve an inversion of the external coordinates, we will refer to them generically as \emph{I-folds}.

We will now review how this happens for the three components of known nine-dimensional theories:\begin{itemize}

\item \textbf{Rank 17}. There is a single component of the moduli space \cite{Narain:1986am,Aharony:2007du}, corresponding to compactification either heterotic string or type I on a circle. We will choose the description in the type I frame.

In this theory, the nine-dimensional I-fold we constructed can be resolved geometrically: A singularity $\mathbb{R}^3/\mathbb{Z}_2$ uplifts to a quotient $(\mathbb{R}^3\times S^1)/\mathbb{Z}_2$ in the 10d type I, with the $\mathbb{Z}_2$ reflecting each coordinate. This configuration has two fixed points, which are both locally of the form $\mathbb{R}^4/\mathbb{Z}_2$. So the 9d I-fold uplifts to two 10d orientifolds of codimension four in the 10d perspective. 

The I-fold includes a nontrivial action on the internal gauge fields of the theory. This means that when parallel-transporting fields along the torsion 1-cycle of the $\mathbb{RP}^2$ transverse to the I-fold, the fields pick up an action of charge conjugation. This 1-cycle is homologous to either of the 1-cycles in the asymptotic ALE geometry of the $\mathbb{R}^4/\mathbb{Z}_2$ singularities, so the same\footnote{In general, we could also have a Wilson line in the $S^1$ before the quotient, in which case the two actions wouldn't need to be the same. But the resulting Wilson line would break the $\text{Spin}(32)/\mathbb{Z}_2$ gauge group far away from the I-fold, too.} internal action in the 9d I-fold must be implemented on each of the $\mathbb{R}^4/\mathbb{Z}_2$ geometries. For the ${\sf CP}$ I-fold, which is defined everywhere on moduli space, the internal action is just ``charge conjugation'', a $\mathbb{Z}_2$ which flips the sign of all the elements in the Cartan, and consequently sends each root to minus itself. 

This gauge action must be built-in in the orientifolds that make up the I-fold. We consider here two kinds of codimension-4 orientifold planes in type I/II string theory, the $O5^+$ and $O5^-$ \cite{Polchinski:1996ry}. They look the same geometrically, but they differ on their action on the 2-form and the open string modes. In particular, the $O5^+$ includes an action on the gauge bundle, implemented by conjugation by the symmetric matrix \cite{Polchinski:1996ry}
\begin{equation} \left(\begin{array}{cc} \mathbf{I}_{16}&0\\0&-\mathbf{I}_{16}\end{array}\right)\label{err}\end{equation}
on the Chan-Paton factors, which live in the adjoint of $SO(32)$. This is precisely the charge conjugation operator ${\sf C}$ defined above, after a gauge transformation to put it in the Cartan algebra (and have it represented by a diagonal matrix).  To see this, notice that the matrix representing any element of the gauge group in the vector representation can be diagonalized by an $SO(32)$ transformation. $\mathbb{Z}_2$ transformations can only have eigenvalues $\pm 1$; and ${\sf C}$ acts on the weights of the vector representation by reversing the sign so that all the states group in pairs $\vert v\rangle ,\vert-v\rangle$ mapped to each other by ${\sf C}$; there are 16 such blocks, and each has eigenvalues $\pm1$, so ${\sf C}$ indeed corresponds to \eq{err}.

This is enough to identify the I-fold with the $O5^+$ plane. We can now consider the compactification  with 8 I-folds on $T^3/\mathbb{Z}_2$, which is just an orbifold K3 $T^4/\mathbb{Z}_2$ with 16 $O5^+$'s from the 10-dimensional perspective. There is an interesting complication: since the charge of the $O5^+$ is positive\footnote{It is $+1/2$ in units where the charge of a fundamental type I $D5$ brane is $+1$. This is consistent with the fact that the type II $O5^+$ has unit type II $D5$ charge \cite{Bergman:2001rp} and also with the Dirac quantization argument we give later in this section, because the dual object is now a D1 string, which has local gravitational anomalies. The gravitational charge of the orbifold is $-3/2$, thus providing the correct shift in the phase. No such phenomenon happens for type II strings.}, the only way to cancel the tadpole is to break supersymmetry, by introducing 8 mobile $\overline{D5}$ branes\footnote{It is also possible to introduce ``half $\overline{D5}$ '' branes at the fixed points, up to 16 of them. But they are stuck at the fixed points and there are consistency conditions on the arrangement \cite{Berkooz:1996iz}, so we focus on the simpler case where they are all sitting on a single fixed point.}. Even though the I-fold is locally supersymmetric, the global compactification is not. On the particular point where all the $\overline{D5}$ are on top of the same orientifold, the perturbative 6d gauge group is $SO(16)^2\times (\text{Sp}(8))^{2}$ \cite{Antoniadis:1999xk,Aldazabal:1999jr,Mourad:2017rrl,Angelantonj:2020pyr}, where the second factor comes from the $\overline{D5}$ branes, and the $SO(16)^2$ factor can be understood from the action of ${\sf CP}$ on the 10d gauge fields, since the group is broken to the commutant of \eq{err}.  More generally, a stack of $n$ $\overline{D5}$ branes has a $\text{Sp}(n)$ symmetry, enhanced to $\text{Sp}(n)^2$ on top of an $O5^+$.

 Even though supersymmetry is broken due to the antibranes, the breaking is mild and localized in a compact space, so the model shares many features with supersymmetric ones. In particular, the sector of the 6d theory coming from 10 supergravity fields is guaranteed to be supersymmetric, since we can compute the dimensional reduction in pure supergravity terms, and so are the multiplets located at the I-folds. The only fields that do not fall in supersymmetric multiplets are localized on the $\overline{D5}$ branes.

Let us study the 10d supergravity fields first. According to our prescription above, the  ${\sf CP}$-even 10d vectors produce 240 6d vectors furnishing the adjoint of $SO(16)^2$, while the 256 ${\sf CP}$-odd vectors produce hypermultiplets(in the $(\textbf{16},\textbf{16})$). The point of the  ${\sf CP}$ action is that it works at a generic point of the 9d Coulomb branch, so it should be possible to turn on 9d Wilson lines. These correspond to a vev of the $(\textbf{16},\textbf{16})$ fields, which can break the gauge group completely. After Higgsing, there are $256-240=16$ hypermultipletsleft, precisely what one would obtain directly from the 9d ${\sf CP}$ action.

As explained in \cite{Polchinski:1996ry}, the $O5^+$ contribute one tensor multiplet each. And finally, $\overline{D5}$ contribute a nonsupersymmetric spectrum that does not arrange itself into hypermultiplets. For details, see \cite{Antoniadis:1999xk,Angelantonj:2020pyr}.

The resulting spectrum should of course be anomaly free; let us look in particular at gravitational anomaly cancellation. A first guess would be that, since the model is non-supersymmetric, the usual anomaly equation
\begin{equation} 273= 29\,n_T+n_H-n_V\label{00000}\end{equation}
needs to be modified. However, there is still a gravitino contributing to the anomaly\footnote{One might wonder how can there be a massless gravitino if supersymmetry is broken. The answer is that the antibranes introduce a dilaton tadpole \cite{Antoniadis:1999xk}, and so flat-space representation theory is not relevant. Having chiral gravitini in a nonsupersymmetric model with a tadpole is a generic situation: for instance, compactify any 10d string on the connected sum of $n$ copies of $K3$. The index theorem predicts $n$ 6d ``gravitini'' per 10d Majorana-Weyl graviton, but this is fine since the background does not have a Ricci-flat metric and is necessarily dynamical.}, and as we have seen all fields except those coming from the $\overline{D5}$ are supersymmetric. Furthermore, in six dimensions, the smallest possible contribution to the $\text{tr}(R^4)$ term (from which \eq{00000} comes from)  is that of a single Weyl fermion of either chirality, i.e. that of a hypermultiplet or a vector multiplet. Self-dual or anti self-dual tensors contribute 28 times this\footnote{The familiar contribution of a tensor multiplet of $29$ splits as $28$ coming from the self-dual tensor and 1 coming from the fermion in the multiplet.}. As a result, whatever the contribution of the antibranes is, we can model it by an effective contribution to $n_H-n_V$, which will be an integer for each antibrane, and still use \eq{00000}. 

In this particular case and on a generic point in the Coulomb branch, we have\footnote{A stack of $8$ $\overline{D5}$ contributes a gauge group $Sp(8)$, enhanced to $Sp(8)^2$ if the branes sit on top of one of the orientifolds. Here, the $\overline{D5}$ contributes  $(\textbf{120},1)\oplus (1,\textbf{120})$ left-handed fermions, as well as 16 copies of $(\textbf{16},1)\oplus (1,\textbf{16})$. The right-handed fermions live in a $(\textbf{16},\textbf{16})$. The net spectrum is that of -240 left-moving fermions.}
\begin{equation}n_T=17,\quad n_H=20,\quad (n_H-n_V)^{\overline{D5}}=-240,\end{equation}
and so \eq{00000} is satisfied. Notice that the contribution of the $\overline{D5}$ branes, as well as the tensors from the $O5^+$, are all multiples of 8. This is a direct reflection of the fact that we need eight identical copies of the same objects to cancel the tadpole. Then, anomaly cancellation would force the contribution of the hypermultipletscoming from the 10d gauge fields to be a multiple of 8 plus 1. Anomalies only cancel precisely because the rank of the 10d/9d gauge group is 1 modulo 8. This is an extremely roundabout way to arrive at something we already knew, but the point is to illustrate the general result which we will derive later in the specific case of string constructions.

Even though we have only constructed the ${\sf CP}$ I-fold in a corner of the moduli space, this is enough to show that the rank of the 9d gauge group is 17 everywhere in that connected component of the moduli space, since it cannot jump. However, it is useful to have a description of the ${\sf CP}$ I-fold directly in other duality frames. In particular, a description in the $E_8\times E_8$ heterotic frame will allow us to generalize the construction directly to the rank 9 component. 

In the heterotic frame, the $\text{Spin}(32)/\mathbb{Z}_2$ I-fold can be obtained from the type I construction by S-duality. The $\overline{D5}$ branes become $\overline{NS5}$-branes, and the $O5^+$ become $\mathbb{R}^4/\mathbb{Z}_2$ with a singular instanton inside, of instanton number $2$ as fixed by the orientifold charges. This shifts the background contribution to the RR charge from $-3/2$ (from the Pontryagin class of the singular bundle) to $+1/2$, as corresponds to the S-dual of the $O5^+$. The monodromy at infinity must be the same as in the type I setup, that is, \eq{err}. As explained in \cite{Berkooz:1996iz}, this is precisely what one gets for a heterotic $\text{Spin}(32)/\mathbb{Z}_2$ instanton with vector structure and instanton number 2. The preserved gauge group is locally $SO(16)^2$. We need to explain the fate of the tensor multiplet in the $O5^+$. The dimension of the instanton moduli space in this singular space is \cite{Berkooz:1996iz}
\begin{equation} \text{dim} \mathcal{M}_2=240-(496-240)/2=112.\end{equation}
This is arranged in terms of $112/4=28$ hypermultiplets. There is an additional hypermultiplet coming from the blow-up mode, which is absent in the type I picture but present in heterotic, for a total of $29$ multiplets. This explains what happens to the type I tensor: it transitions to 29 hypers, as in the heterotic small instanton transition.

Having understood the $\text{Spin}(32)/\mathbb{Z}_2$ I-fold, the $E_8\times E_8$ can be obtained by turning on a Wilson line along one of the directions of the parent $T^4$, T-dualizing, and subsequently switching off the line. Alternatively, one can construct the ${\sf CP}$ action directly, using the results of \cite{Berkooz:1996iz} as in the  $\text{Spin}(32)/\mathbb{Z}_2$ case. The I-fold is again a $\mathbb{R}^4/\mathbb{Z}_2$ singularity with one instanton in each $E_8$. Compactification again requires the presence of 8 pointlike $\overline{NS 5}$ branes. 

\item \textbf{Rank 9 theories}. Starting with the $E_8\times E_8$ in the presence of I-fold, constructing the CHL version is straightforward, as the requisite $\mathbb{Z}_2$ involution commutes with ${\sf CP}$. Identifying the two $E_8$'s reduces the instanton number by half, but the contribution to the fivebrane tadpole is the same. The compactification is again nonsupersymmetric, and still requires 8 $\overline{NS 5}$ branes. 

The model also satisfies gravitational anomaly cancellation. At the $E_8\times E_8$ point, the gauge group is broken to $SO(16)\times SO(16)$, so we get $256$ hypermultipletsfrom the 10d gauge fields. After CHL orbifolding, half of these are projected out, resulting in 128 hypermultipletsonly. Due to the WL that exchanges the two $E_8$'s, half of the 16 fixed points involve an additional action swapping the two $E_8$'s. The associated instanton moduli space has dimension
\begin{equation} \text{dim} \mathcal{M}=240-(248-128)/2=176.\end{equation}
This means that each of the twisted fixed points contribute $176/4=44$ hypermultiplets, instead of 28. So there is 16 additional hypermultipletsper twisted fixed point. Their net additional contribution is $16\cdot 8=128$, so the anomaly again cancels. 

\item \textbf{Rank 1 theories.} We just need to construct the I-fold in the rank 1 theories to show that it exists in every known string example. This case is particularly simple since there is no internal gauge group, and so the ${\sf CP}$ and ${\sf P}$ actions are equivalent. It is also supersymmetric, in contrast with previous examples.  For the branch of moduli space realized as M theory on the Klein Bottle, the I-fold is simply 
\begin{equation}(T^2\times \mathbb{R}^3)/(\mathbb{Z}_2\times\mathbb{Z}_2),\label{KBac}\end{equation}
 where the first $\mathbb{Z}_2$ acts on the $T^2$ coordinates $(x,y)$ (of periodicity 1 each) as
 \begin{equation}(x,y)\,\rightarrow\, \left(x+\frac12,-y\right)\label{kbac1}\end{equation}
  to produce the Klein bottle, and the second action acts on $y$ as well as the $\mathbb{R}^3$ vector $\vec{z}$ as
  \begin{equation} (x,y,\vec{z})\,\rightarrow\, \left(x,-y,-\vec{z}\right).\end{equation}
  Both actions commute. The second has two fixed lines, each of which is locally of the form $\mathbb{R}^4/\mathbb{Z}_2$ (so just $A_1$ singularities). Quotienting by \eq{kbac1}, this introduces two fixed points, where the local geometry is of the form $\mathbb{R}^5/\mathbb{Z}_2$.   At these points we have $MO5$ planes. These can carry half-integer quantized M5 brane charge \cite{Dasgupta:1995zm,Witten:1995em}.

  We can repeat the above on the compact space $T^5/(\mathbb{Z}_2\times \mathbb{Z}_2)$.  It has eight fixed lines $\mathbb{R}^4/\mathbb{Z}_2$, as well sixteen fixed points, with a tadpole cancelled by 8 M5 branes. The $\mathbb{R}^4/\mathbb{Z}_2$ can be regarded as IIA orbifolds, each of which contributes one vector and one hypermultiplet. The M5 branes are locally $(2,0)$ supersymmetric; the worldvolume degrees of freedom live in a $(2,0)$ tensor multiplet. This decomposes as a tensor plus a hypermultiplet in $(1,0)$ language \cite{polchinski1998string}. In fact, the model we have is exactly the one considered in \cite{Sen:1996tz}, where the $K3$ is in the orbifold limit. The 9d supergravity fields lead to the usual  gravity multiplet, tensor, and four hypers. The resulting spectrum has
  \begin{equation} n_T=9,\quad n_H=20,\quad n_V=8,\label{anompark}\end{equation}
  and so anomalies cancel.
 
 Finally, we just need to construct the I-fold in the Dabholkar-Park background \cite{Dabholkar:1996pc}. This is most conveniently done in the IIA frame where this branch of the moduli space can be described as an $O8^+-O8^-$ compactification. The I-fold is simply an $O6$ plane stretching between the two $O8$'s. This theory can be also described as a $\mathbb{Z}_2$ orbifold of IIB involving a shift of a coordinate as well as orientation reversal. This shift commutes with the purely geometric $\mathbb{Z}_2$ action on $T^4$ that leads to an orbifold K3, so we can consider the Dabholkar Park orientifold of an orbifold K3. This is precisely the I-fold compactification. The Wilson line means that we have 8 points which are pure IIB orbifolds $\mathbb{R}^4/\mathbb{Z}_2$, while the other eight are fixed by $\Omega$ and reflection of four coordinates. These eight fixed points have no localized degrees of freedom, but contribute $-1$ to the RR tadpole \cite{Polchinski:1996ry}. We then need to introduce 8 $D5$ branes to cancel the tadpole. The worldvolume of the $D5$'s is locally $(1,1)$ supersymmetric, and the vector multiplet becomes a vector and a hypermultiplet in $(1,0)$ language. 
On the other hand, the type IIB string on $\mathbb{R}^4/\mathbb{Z}_2$ leads to the $(2,0)$ theory with a single tensor from the dimensional reduction of the self-dual 4-form, so \eq{anompark} applies again. Thus we end up with an anomaly-free 6d model with $U(1)^8$ gauge symmetry, in fact discussed in \cite{Dabholkar:1996zi} (see also \cite{Polchinski:1996ry}), so that \eq{anompark} applies again. The I-fold compactifications of the two 9d rank 1 models should be related to each other, since the two models are equivalent under compactification to 8 dimensions, where they are T-dual \cite{Dabholkar:1996pc,Aharony:2007du}. And indeed, the corresponding 6d I-fold backgrounds are dual to each other \cite{Dabholkar:1996zi}.
\end{itemize}

The fact that I-folds exist in all known 9d compactifications (and hence, also in all 8d examples, by dimensional reduction) is a nice consistency check of the cobordism conjecture, which predicts their existence, which is unavoidable if one wants to evade a global $\mathbb{Z}_8$ symmetry (the other option, that parity is not gauged, would have meant that the theory cannot be put on $\text{Pin}^-$ manifolds in the first place). Therefore, they must exist too in any other would-be 9d theory of gravity we may not know about. As we will see, if the ranks of the gauge groups in 9d are not equal to 1 mod 8, this leads to an anomaly in the lower dimensional theory, and therefore puts them in the Swampland. 

Another interesting feature is that the local I-folds constructed above in stringy examples are all BPS, preserving 8 supercharges (we emphasize however that this is not necessary for our arguments). A geometric action like \eq{geom-orb} is certainly compatible with supersymmetry, so it is natural to expect that the resulting I-folds are BPS in a general $\mathcal{N}=1$ theory.  One could perhaps envision some form of the completeness principle \cite{Polchinski:2003bq} that ensures that all representations of the relevant supergroup are present in a consistent quantum theory of gravity\footnote{A version of completeness involving spacetime symmetries is natural. Not only there should be physical states on every representation of the internal symmetry group, but also in every representation of SUSY (and hence Poincar\'{e}) groups. The Poincar\'{e} part is trivially satisfied in any Einstein quantum theory of gravity, since multi-particle graviton states can have any mass and integer total spin (by coupling internal and external spin). It would be interesting to explore this further.}, which could perhaps explain the BPS character of the I-folds.  

Even though the I-folds themselves are supersymmetric, I-fold compactifications in the rank 17 and rank 9 cases are not. This is because the I-fold carries additional conserved charges that must be cancelled, and the requisite objects break supersymmetry.

In any case, we do not actually need the compactification or even localized degrees of freedom at the I-fold to be supersymmetric. All that we need is that the boundary conditions that the I-fold sets for the supergravity fields preserve supersymmetry; these are in turn fully specified by the requirement that the I-fold is a nulbordism for $\mathbb{RP}^2$. The reason why we can be so cavalier about supersymmetry is the usual fact that anomalies are quite robust due to their topological character, as explained around equation \eq{00000}.  Indeed it is quite satisfying to see that we consider supersymmetry breaking backgrounds to find restrictions on the rank of gauge groups in supersymmetric theories!

\subsubsection{I-fold compactifications and charge quantization}\label{sec:tach}

We will now see how much of the features of the above examples can be argued in general. We will be considering toroidal $T^3/\mathbb{Z}_2$ compactifications with I5-folds at the fixed points. As the examples above illustrate, the Ifolds can carry additional conserved charges, which can lead to tadpoles. These must be cancelled by introducing appropriate sources, like the $\overline{D5}$ branes above,  and in our arguments below it will be important to know exactly how many fivebranes of this kind we need to introduce. To do this we now discuss the possible charges that an I5 can carry, as well as their quantization properties.

As discussed above, the I5-fold carries a $\mathbb{Z}_8$ gauge charge, which is measured by the fact that its asymptotic geometry is $\mathbb{RP}^2$. There is an associated  $\mathbb{Z}_8$ tadpole, but it is automatically cancelled in the $T^3/\mathbb{Z}_2$ background since it has 8 fixed points.

However, as the stringy examples show, this is not the only charge that the I5 can carry. In nine dimensions, it has the same dimensionality as a magnetic 5-brane for a $U(1)$ gauge field. A magnetic 5-brane of positive charge preserves the same supercharges as the I5, thus, it is possible to stick magnetic charge on top of it. The magnetic charge is measured by the topology of the $U(1)$ bundle in the asymptotic $\mathbb{RP}^2$; because the reflection symmetry we have reverses gauge fields, the relevant cohomology group has twisted coefficients, $H^2(\mathbb{RP}^2,\tilde{\mathbb{Z}})=\mathbb{Z}$ \cite{Bergman:2001rp} and hence we can put a gauge flux on top of the I5. 

On a compact space, like  $T^3/\mathbb{Z}_2$, the total magnetic charge must vanish. Typically, we will need to put additional branes to cancel the tadpole. How many will depend on what is the magnetic charge that the I-fold can carry. Since it is not an asymptotically flat geometry, the I-fold could carry fractional charge in principle.

 We can work out the charge quantization of the I-fold by a generalization of the Dirac quantization argument for orientifold backgrounds, explained e.g. in \cite{Tachikawa:2018njr}. One usually determines the Dirac quantization condition by computing the Aharonov-Bohm phase picked up by an electrically charged particle as it moves in the field of the magnetic source. When the angular geometry around the monopole is not a sphere, as it happens for the I5-fold, the worldvolume fields of the electrically charged object could contribute an additional term to the Dirac quantization phase. For the I5 in nine dimensions, the electric object is a particle, and a single Majorana $\text{Pin}^-$ worldvolume fermion would add contribution of 
 \begin{equation}\eta(\mathbb{RP}^2)=1/8\label{rpfer2}\end{equation}
 to the Dirac phase, setting the Dirac quantization condition of the I-fold to $1/8$ the charge quantum of an ordinary magnetic 5-brane. This would jeopardize our arguments below, and as we saw above it is not the behavior found in string compactifications. 

The day is saved by the fact that we are considering a theory with sixteen supercharges. These must be represented in the 1d worldvolume of the electrically charged particle, as ``goldstino'' operators. There must be sixteen fermion zero modes  for a non-BPS particle and eight of them for a BPS one.  Each fermion zero mode contributes to the Dirac pairing the same as a single $\text{Pin}^-$ worldvolume fermion, that is as in \eq{rpfer2}. Since the contribution from eight fermions vanishes, we conclude that the I5 carries integer-quantized monopole charge. There is a subtlety that needs to be addressed in this argument, since one could imagine there could be additional internal degrees of freedom that change the contribution to the parity anomaly. This is not possible in the present case, due to supersymmetry, as discussed in Appendix \ref{app:parity1d}.

 The phenomenon of non-integer quantized charges on backgrounds with nontrivial angular geometry is familiar in orientifolds \cite{Bergman:2001rp,Tachikawa:2018njr}. For completeness, we now discuss an example (taken from \cite{Tachikawa:2018njr}) where supersymmetry is not enough to conclude integer quantized charges. The setup is the IIA $O4^+$ plane, for which the electric probe is a BPS $D2$ brane. Briefly, the fact that the angular geometry for the $O4^+$ is $\mathbb{RP}^4$ means that a $\text{Pin}^+$ worldvolume fermion in the $D2$ brane will add a contribution of $\eta(\mathbb{RP}^4)=1/16$ to the Dirac pairing\footnote{In general, there could be other contributions to the worldvolume anomaly. For instance, an $O3^-$ induces a nontrivial $SL(2,\mathbb{Z})$ bundle in the angular $\mathbb{RP}^5$, and the worldvolume gauge fields of the probe $D3$ become sensitive to this \cite{Hsieh:2019iba}. Interestingly, this kind of effect only happens for $Op^-$ planes; for $Op^+$, only the fields in \cite{Tachikawa:2018njr} contribute. In our case, the $I5$ does not source any other fields in the low-energy supergravity that could contribute to the anomaly; this is also what happens for e.g. $O4$ planes, which have half-integer quantized charges due to the contribution of worldvolume fermions but no additional contribution from the worldvolume gauge fields.}.  Since the $D2$ has $\mathcal{N}=8$ supersymmetry in three dimensions, there are eight gauginos, and their total contribution to the Dirac pairing is a sign, which offsets the contribution coming from the half-integer quantized charge of the $O4^+$. This is the correct quantization condition. 

This example uplifts to a $T^5/\mathbb{Z}_2$ orientifold in M theory, with 32 fixed points, discussed in \cite{Dasgupta:1995zm,Witten:1995em}. As explained there, the $MO5$'s carries $-1/2$ units of discrete four-form flux. Due to this, one only needs $16$ M5 branes to cancel the tadpole (instead of 32).  There are several ways to cancel the tadpole, but none in which the 32 fixed points are treated symmetrically. This matches with the I-fold for M-theory in the Klein bottle we found in Section \ref{sec:Ifolds}, where the number of fixed points is reduced from 32 to 16 and the number of $M5$-branes from 16 to 8. A 9d I5-fold is composed of two M-theory $MO5$-planes, so its charge is integer quantized in 9d,  consistently with our argument.

Another way of understanding the difference between M-theory and our 9 dimensional theories comes from looking at the bordism class that the orientifold is killing. In M-theory, the $MO5$ is killing the class of $\mathbb{RP}^4$; as remarked above, M-theory makes sense on $\text{Pin}^+$ manifolds, but when the fourth Stiefel-Whitney class $w_4$ is nontrivial, it actually requires to turn on a half-integer quantized $G_4$ flux \cite{Witten:1996hc}. So the $\mathbb{RP}^4$ comes with $G_4$ flux attached, which must be eaten up by the orientifold, thereby leading to half-integer quantized M5 charge. By contrast, there is no such constraint in the nine-dimensional theory; the only tadpole condition is the Bianchi identity for the $B$-field, which is not relevant for $\mathbb{RP}^2$. 

These are all the charges relevant\footnote{The I5 plane can also carry a holonomy of the antisymmetric $B$ field, corresponding to $H^2(\mathbb{RP}^2,\mathbb{Z})=\mathbb{Z}_2$. This implies the existence of $I5^+$ and $I5^-$ in general, just like $O5^\pm$ in known string theories. Notice that this holonomy does not lead to a tadpole since $\int_{\mathbb{RP}^2} B_2$ is not a bordism invariant. Similarly, one can turn on a $\mathbb{Z}_2$ Wilson line corresponding to $H^1(\mathbb{RP}^2,\tilde{\mathbb{Z}})=\mathbb{Z}_2$; again, there is no associated bordism invariant so it leads to no new tadpoles.} to the I-fold compactification\footnote{The above are the charges that can be measured using the supergravity fields. We will assume there are no other conserved charges that lead to additional tadpoles. For instance, there could have been some hidden $\mathbb{Z}_{16}$ gauge symmetry which is completely invisible from the supergravity point of view. Cancelling tadpoles for this charge would require different kinds of $I5$-folds in the fixed points of $T^3/\mathbb{Z}_2$, which would change our conclusion below.  Again, these charges do not exist in stringy examples, where explicit compactification is possible. For now,  will just assume no additional charges exist.}. So to summarize, because of the above, we will assume that in any consistent $\mathcal{N}=1$  supersymmetric theory, the orbifold $T^3/\mathbb{Z}_2$ is an allowed supersymmetric background of the theory (at least in the supergravity sector); the eight fixed points cancel the geometric $\mathbb{Z}_8$ charge of the I-fold, and due to the Dirac quantization argument above, any monopole charge they generate can be cancelled by introducing an integer number of monopoles  of unit charge per fixed point. These monopoles in turn exist due to the completeness hypothesis.

We will now explore the consequences of the existence of the I-fold compactifications in 9, 8 and 7 dimensions.

\subsection{Rank 8 periodicity in \texorpdfstring{$d=9$}{d=9}}\label{sec:9dmod8}

As explained above, compactification of the 9d theory on $T^3/\mathbb{Z}_2$ leads to a 6d $(1,0)$ supergravity theory since the geometric action background projects out eight supercharges. The supersymmetry might be broken by the localized defects, as in the explicit stringy I-folds in Section \ref{sec:Ifolds}. Massless states can arise either from dimensional reduction of 9d supergravity states or as twisted states localized at the $I5$'s, as follows:\begin{itemize}
\item Dimensional reduction of the bosonic sector of the gravity multiplet on $T^3$ leads to a 6d graviton, two-form (without a duality constraint), and vector, three Kaluza-Klein vectors, three ``winding'' vectors coming from the 2-form with one index in the internal direction, and thirteen scalars (the 9d scalar plus three Wilson lines for the vector, plus six moduli from $T^3$, plus the periods of the 2-form). The geometric action projects out the KK and winding vectors.  We conclude that the massless spectrum is the gravity multiplet, one tensor, and three hypers.   More details and the reduction of the fermions can be found in Appendix \ref{app:dimred}.

\item As discussed in Appendix \ref{app:dimred}, there are two kinds of vector multiplets, depending on whether the scalar is even (and the vector odd) or odd (and the vector even) under parity. The first kind leads to a hypermultiplet, and the second kind, to a vector multiplet. As discussed in Section \ref{sub:par}, under ${\sf CP}$ all vector multiplet scalars are even, so in this case we end up with $r$ hypers. 
\end{itemize}
On top of this, we have to include the contributions from the twisted sectors living on the eight $I5$'s. As has been argued above, there are no discrete charges on the I5 planes, and there are eight identical copies of them. As a result, whatever amount of tensors, vector and hypermultipletsis contributed by the I5's, it has to be a multiple of eight. There will also be a contribution coming from the (possibly supersymmetry breaking) additional branes introduced to cancel the tadpole.  Similarly to the discussion in Section \ref{sec:Ifolds}, all that can happen is that the chiral matter is not arranged in supermultiplets, and instead we get an arbitrary number (possibly negative) of anti-self-dual tensors and Weyl fermions. Because the anomaly of the anti-self-dual tensor is an integer multiple of that of the Weyl fermion, they contribute an effective number of nonsupersymmetric hypermultiplets$n_H^{\text{Nonsusy}}$ to the gravitational anomaly.    Since as we have already argued the number of 5-branes is a multiple of 8, their contribution to anomalies will also be a multiple of 8. 
Putting everything together, we have a 6d  theory with
\begin{equation} n_T=1+8\, n_T^{I5},\quad n_V=8\, n_V^{I5},\quad n_H=3+r+8\,n_H^{I5}+8\,n_H^{\text{Nonsusy}} .\end{equation}
The 6d anomaly cancellation condition (see e.g. \cite{Kumar:2010ru,Taylor:2019ots}) 
\begin{equation}273=29\,n_T+n_H-n_V\label{anocanc6dg}\end{equation}
then implies that 
\begin{equation}241=8\,(29\,n_T^{I5}+n_H^{I5}-n_V^{I5}+n_H^{\text{Nonsusy}})+ r,\label{VVV}\end{equation}
which becomes
\begin{equation}r\equiv 1\,\text{mod}\,8,\label{VVVVVV}\end{equation}
as advertised. 

Thus, the existence of I-folds, or equivalently, completeness together with demanding that the 9d theories make sense in non-orientable manifolds is sufficient to explain the observed periodicity of the rank of the gauge group in nine dimensions. 

Before moving on, it is worth considering why we needed to look at I-folds in the first place. If the 9d theory makes sense on $\text{Pin}^-$ manifolds, why not just look at ordinary parity anomaly directly in 9 dimensions? The relevant bordism group controlling the anomaly theory is $\Omega_{10}^{\text{Pin}^-}$, which is way too large; in particular, it contains a $\mathbb{Z}_{128}$ factor, generated by $\mathbb{RP}^{10}$, which would lead to a mod 128 periodicity in the rank. The caveat is that there can be tadpoles that can prevent us from considering certain backgrounds.\footnote{For instance, one can show that, for M theory on the Klein bottle, it is not consistent to evaluate the anomaly for the rank 2 theory on $\mathbb{RP}^{10}$. This uplifts to evaluating the 12-dimensional anomaly theory on some 12-dimensional manifold $X_{12}=(\mathbb{S}^{10}\times T^2)/\mathbb{Z}_2$, where the $\mathbb{Z}_2$ acts by an antipodal mapping on $S^{10}$ and simultaneously reversing  and shifting the fibers of $T^2$. This is an unorientable $\text{Pin}^+$ manifold, as required in M-theory, but it is not an $\mathfrak{m}^c$ manifold \cite{Witten:1996hc}; the fourth Stiefel-Whitney class has no $w_1$-twisted integer lift. M-theory is not consistent on such backgrounds without additional ingredients.  Basically, the Chern-Simons terms in M-theory prevent us from considering the anomaly theory in the above background. It would be very interesting to understand this phenomenon in generality.} That being said, we note that compactification of the 9d theory on a manifold of $\text{Spin}(7)$ holonomy with $p_1=0$ produces a 1d theory with one supercharge. There is a $\mathbb{Z}_8$ parity anomaly controlled by $\Omega_{8}^{\text{Pin}^-}=\mathbb{Z}_8$, as well as a tadpole that must be cancelled by the introduction of electrically charged particles; as we discussed above, these have no net contribution to the anomaly. Assuming no further obstructions to compactification on this background, the anomaly also seems to capture the modulo 8 periodicity of the rank.

\subsection{Modulo 8 periodicity in \texorpdfstring{$d=8$}{d=8}}\label{sec:8dmod8}

Now, let us consider the 8d case.
Again, we can compactify on $T^3/\mathbb{Z}_2$, using the ${\sf\tilde{C}P}$ symmetry introduced in Section \ref{sec:par8d0}  for the orbifold. The fixed loci are I4 planes this time, and the previous analysis goes through in a similar fashion. We end up with a five-dimensional $\mathcal{N}=1$ theory. There are no local anomalies in five dimensions, so it might seem that we are at an impasse. However, since ${\sf \tilde{C}P}$ commutes with ${\sf P}$, this last 8d parity symmetry becomes a good 5d parity symmetry, so that we can put the theory on $\text{Pin}^-$ manifolds. As always in odd dimensions, there are two $\text{Pin}^-$ representations, which differ by a sign in the product of reflections on all space and the time coordinate.  Because
\begin{equation}\Omega_{6}^{\text{Pin}^-}=\mathbb{Z}_{16},\label{q21}\end{equation}
which is generated by $\mathbb{RP}^6$, there is a potential anomaly. To study this anomaly, we need to know the parity properties of the 5d $\mathcal{N}=1$ multiplets. There is a unique choice for the parity action on gravity multiplet and hypermultiplet, but just as in nine dimensions, there are two different vector multiplets, distinguished by how the scalar transforms under parity. They contribute with opposite signs to the anomaly measured by \eq{q21}. Just as in nine dimensions, we will call them ``vector'' (if the scalar is parity-odd, and the vector parity-even) and ``pseudovector'' (if the scalar is parity-even, and the vector is parity-odd). 

How many multiplets of each kind arise in the quotient theory under ${\sf P}$? An easy way to determine this is to put the 6d $(1,0)$ theory on a circle, and use the realization of ${\sf P}$ as a combined reflection on the circle and the noncompact directions. In six dimensions on a $\text{Spin}$ manifold, the gravitino and hypermultiplet fermions have one chirality, and the fermions in the vector multiplets have the opposite. Upon dimensional reduction these two chiralities correspond to the two different representations of $\text{Pin}^-$. Dimensional reduction of a 6d vector produces a parity-odd scalar (since it is odd under reflection of the internal coordinate), and thus a vector.  On the other hand, a 6d tensor multiplet has fermions of opposite chirality to those of the vector, and there is a scalar in the multiplet. When reducing to 5d, it produces a pseudovector multiplet, with the opposite parity transformation for the fermions. Finally, the 6d hypermultiplet becomes a 5d hypermultiplet and the 6d gravity multiplet becomes the 5d gravity multiplet plus a pseudovector (coming from the 6d anti-self-dual tensor and whose scalar is the radion of the $S^1$).

Since the 9d gravity multiplet becomes the 8d gravity multiplet plus a vector and similarly a 9d vector becomes an 8d vector, we can determine the supergravity spectrum of the 8d theory on $T^3/\mathbb{Z}_2$ by taking the relevant results from 9d and reducing on a circle. One obtains that the 8d gravity multiplet yields then a gravity multiplet, two pseudovectors and two hypermultiplets(the 3 hypermultipletscoming from the 9d gravity multiplet minus one hypermultiplet from the vector that arises in the reduction), while the 8d vector yields a single hyper. In total, we have a 5d theory with $r$ hypers, two pseudovectors, and a gravity multiplet. 

The only contributions to the anomaly come from the fermions in the different multiplets via their $\eta$ invariant \cite{Witten:2015aba,Yonekura:2016wuc}; for the standard Dirac operator in either of these representations, $\eta(\mathbb{RP}^6)=\pm1/16$ \cite{GilkeyBook}, where the sign depends on the $\text{Pin}^-$ representation of the fermion. The anomaly theory of the gravitino on a six-manifold $X$ is \cite{Hsieh:2020jpj} 
\begin{equation}\eta^{\text{gravitino}}=\eta(\mathcal{D}^{\text{Dirac}\times TX})-2\eta(\mathcal{D}^{\text{Dirac}}).\end{equation}
$\eta(\mathcal{D}^{\text{Dirac}})$ is the anomaly of one Dirac fermion with the same parity transformation properties, which we computed before. We just need to compute $\eta(\mathcal{D}^{\text{Dirac}\times TX})$, which can be done as follows: First, notice that $T(\mathbb{RP}^6)\oplus\mathbb{R}\sim 7\mathcal{L}$, where $\mathcal{L}$ is the canonical line bundle on $\mathbb{RP}^6$. So  on the one hand we have $\eta(\mathbb{RP}^6)=7\eta(\mathcal{L})-\eta$, where $\eta=\eta(\mathcal{D}^{\text{Dirac}})$. On the other hand, $\eta(\mathcal{L})$ is just the eta invariant of a fermion coupled to an additional $\mathbb{Z}_2$ bundle; since $\text{Pin}^-$ structures are parametrized affinely by $\mathbb{Z}_2$ bundles, it follows that $\eta(\mathcal{L})$ is just the eta invariant of the fermion with the opposite $\text{Pin}^-$ structure on $\mathbb{RP}^6$. This allows one to evaluate $\eta(\mathcal{L})=-\eta$, and thus, $\eta(\mathbb{RP}^6)=-8\eta$. As a result, the gravitino contributes as ten Dirac fermions. Taking into account its chirality produces an overall minus sign. Including parity transformation, we obtain the following contributions to the anomaly:

\begin{equation}
\begin{array}{c|cccc}
\text{Multiplet}&\text{Gravity}&\text{Hyper}&\text{Vector}&\text{Pseudovector}\\\hline
\text{Anomaly}&5/8&1/16&-1/16&1/16
\end{array}
\end{equation}

As before, the total anomaly polynomial of the fermions of an 8d theory of rank $r$ on a $T^3/\mathbb{Z}_2$ I-fold is that coming from the supergravity field plus an undetermined contribution which is a multiple of 8 coming from the I4 planes. As a result, the condition of cancellation of total anomaly for the ${\sf P}$ symmetry is simply
\begin{equation} r+14+8k\equiv 0\,\text{mod}\,16\quad\Rightarrow\quad r\equiv2\,\text{mod}\,8.\label{ss8d}\end{equation}
We therefore obtain the correct periodicity observed in string compactifications to eight dimensions \cite{Kim:2019ths}. 

There is one subtlety we must discuss. Anomaly cancellation in six dimensions is very robust, and there is no way around \eq{anocanc6dg} without introducing additional massless degrees of freedom. The five-dimensional reduced version \eq{ss8d}, however, is just a discrete $\mathbb{Z}_{16}$ anomaly, and anomalies of this kind can sometimes be cancelled by coupling to a topological quantum field theory (tQFT), in what we might call a topological  version of the Green-Schwarz mechanism \cite{Garcia-Etxebarria:2017crf} (see \cite{2015ARCMP...6..299S} for a condensed matter viewpoint). Such a coupling does not alter  the massless content or interactions of the gapless sector of the theory\footnote{This should be done in a supersymmetry-compatible way, which might introduce further restrictions.}. If this were possible, the tQFT could weaken the anomaly cancellation condition \eq{ss8d}, and the precise match with string theory would be lost.   

Before discussing this possibility in some detail, we now argue that complete anomaly cancellation via this mechanism is not possible in this case.
Using the internal ${\sf \tilde{C}}$ symmetry in eight dimensions,  the cobordism conjecture predicts that we can compactify instead on $T^2/\mathbb{Z}_2$ in such a way that the resulting 6d $(1,0)$ theory imposes, again via local gravitational anomaly cancellation, that the rank is $2\,\text{mod}\,4$, so the mod 8 periodicity is at least partially explained. Thus, even if the top. GS mechanism can take place,  \eq{ss8d} is weakened to a $\text{mod}\,4$ condition, but no further. 

The topological GS mechanism is a real possibility, but does it actually happen for $\Omega_6^{\text{Pin}^-}$? The subject of which anomaly theories can be realized by a tQFT was recently analyzed in \cite{Wan:2018zql,Wan:2018djl,Cordova:2019bsd}, where it was shown that if the anomaly theory is nontrivial on mapping tori, then it can never be described by a tQFT. In our case, this would work if we could find a mapping torus bordant to the connected sum of two $\mathbb{RP}^6$; we have not been able to do so.  On the other hand, the symmetry extension technique of \cite{Witten:2016cio,Wang:2017loc,Tachikawa:2017gyf,Kobayashi:2019lep} which allows one to construct anomaly-matching tQFT's in many cases, does not apply straightforwardly to $\Omega_6^{\text{Pin}^-}$\footnote{We thank Kantaro Ohmori for explaining this to us.}. 

So far, we have just described the field theory picture. Since a tQFT is UV-complete on its own, in a field theory setup one can just add it on top of whatever else there is, without any further effects. As usual, things are subtler in quantum gravity. The observables in a tQFT are (extended) line, surface operators etc. whose correlators are topological. This is forbidden by a natural generalization of the completeness principle/absence of global symmetries in quantum gravity \cite{Polchinski:2003bq,Banks:2010zn,Rudelius:2020orz}. It is also a consequence of the no Baby Universe Hypothesis \cite{McNamara:2020uza}, which states that there are no genuine compact operators in quantum gravity.

So one cannot really have a tQFT nontrivially coupled to gravity. However, this does not mean that the anomaly cannot be cancelled by gapped degrees of freedom; there might be very massive local fields that do the trick. At energies much below the gap, the physics of these massive states is described by a ``low-energy effective tQFT''. Such an effective tQFT can arise as part of the low-energy effective action, and it can cancel anomalies such as the one we consider here, even if it we just argued it cannot happen at any energy scale\footnote{Correlators of extended operators are (almost) invariant under deformations at low energy; at high enough energies, we can ``break them open'', but at such energy scales the theory is no longer topological.}. 

As an example of such a low-energy effective tQFT in a concrete string compactification, take the heterotic $T^6/\mathbb{Z}_3$ orbifold discussed in Section 9.5.1 of \cite{Ibanez:2012zz}. At low energies there is a $U(1)$ which becomes massive due to the (ordinary) Green-Schwarz mechanism, and the low-energy effective theory below the string scale includes a topological $BF$ theory at level 9. That being said, the fact that we can find \emph{some} tQFTs at low energies does not mean we can find \emph{any} tQFT. In other words, there is a Swampland of low-energy tQFTs. For instance, a 5d abelian CS theory
\begin{equation} k\int F\wedge F\wedge A\end{equation}
is a perfectly good tQFT, but per the arguments above we expect the Wilson lines to be breakable in the UV; at this scale, there are electrically charged particles, and the gauge field is dynamical. We would expect to have a kinetic term for $A$, but then the theory is no longer topological, since there are massless excitations. Thus, the pure 5d CS theory should be in the Swampland, while BF is not. We leave the interesting question of the Swampland of low-energy tQFT's for future work. 

We also note that supersymmetry must also be taken into account, and this might place further restrictions on the putative tQFT that could cancel the anomaly.

To sum up, we have been unable to prove that the topological Green-Schwarz mechanism cannot happen in our case, but even if it did, the tQFT might turn out to be in the Swampland, or not compatible with supersymmetry.  In any case, the rank periodicity is true at least mod 4. The weaker mod 4 condition would leave the rank 6 and rank 14 possibilities open. It also is conceivable that there are yet to be discovered M/string theory compactifications to 8d with rank 6 and 14.

\subsection{Modulo 2 periodicity in \texorpdfstring{$d=7$}{d=7}}\label{sec:7dmod2}

As explained in Subsection \ref{sec:par7d}, the 7d theory makes sense in either $\text{Pin}^+$ or $\text{Pin}^-$ manifolds. The cobordism conjecture applied to
\begin{equation}\Omega_0^{\text{Pin}}=\mathbb{Z}_2\end{equation}
predicts that there must be a domain wall for this theory, which will naturally be an I6, so we are allowed to consider the theory on\footnote{It is important we assume non-orientability, since that way we conclude that two identical I6-folds cancel the tadpole. Trying to play the same game using $\Omega_0^{\text{Spin}}=\mathbb{Z}$, we would have obtained two potentially different objects at the endpoints of the interval, since a point ``remembers'' its orientation in Spin bordism.} $S^1/\mathbb{Z}_2$. 

The 7d vector multiplet contains now 3 scalars, and the gravity multiplet contains three vectors and a second dilatino. On $S^1/\mathbb{Z}_2$, the gravity multiplet yields a 6d gravity multiplet, a tensor, and a single hypermultiplet. The vector produces one hypermultiplet as usual.  Local anomaly cancellation taking into account that the two I6's produce the same contribution then implies
\begin{equation} r\,\equiv\,1\,\text{mod}\,2\label{7danom44}\end{equation}
as we claimed.

There is a second way we can obtain the same result, which even though more complicated, is in line with our previous considerations. As mentioned in Section \ref{sec:par7d}, there may be points of the moduli space where both $\text{Pin}^+$ and $\text{Pin}^-$ symmetries are present; in these cases, the symmetry group becomes $\text{Pin}^{\mathbb{Z}_4}$.  We can then compactify on $T^3/\mathbb{Z}_2$, which does not break the $\mathbb{Z}_4$ symmetry, and  ask about anomalies under the $\text{Spin}^{\mathbb{Z}_4}$ subgroup. As described in \cite{Tachikawa:2018njr,Garcia-Etxebarria:2018ajm},
\begin{equation}\Omega_{5}^{\text{Spin}^{\mathbb{Z}_4}}=\mathbb{Z}_{16},\end{equation}
so there can be an anomaly.

The 7d gravity multiplet on $T^3/\mathbb{Z}_2$ yields a 4d gravity multiplet, three vectors, and two hypers, while a 7d vector just yields two hypers. In total, we have a 4d gravity multiplet, three vectors, and $2r+2$ hypers. The $\text{Spin}^{\mathbb{Z}_4}$ is an R-symmetry, with the two chiral supercharges (of the same chirality) transforming with a phase. This means that any Weyl fermion in any multiplet transforms as $\pm i$. The $\eta$ invariant of any such fermion is $\pm1/16$ \cite{Garcia-Etxebarria:2018ajm}. A similar calculation to the one carried out in Subsection \ref{sec:8dmod8} shows that the ($\mathcal{N}=1$) gravitino contribution is seven times that of a Dirac fermion (see also \cite{Garcia-Etxebarria:2018ajm}). The contributions of the different multiplets to the anomaly are as follows:

\begin{equation}
\begin{array}{c|ccc}
\text{Multiplet}&\text{Gravity}&\text{Hyper}&\text{Vector}\\\hline
\text{Anomaly}&7/8&1/8&-1/8
\end{array}
\end{equation}
Here, we have taken into account that a hypermmultiplet contains two Weyl fermions, and the gravity multiplet contains two gravitini. Taking into account the full spectrum, we get that the anomaly cancellation condition is
\begin{equation} 7+2(r+1)-3\equiv 4 k \,\text{mod}\, 8,\end{equation}
where the rhs takes into account the contribution of twisted sectors. This is again \eq{7danom44}. Thus, we recover the modulo two periodicity. In this case, we could worry a priori about a topological Green-Schwarz mechanism making the anomaly weaker, but we know this cannot be the case as we already gave an alternative derivation for the mod 2 condition based on 
$S^1/\mathbb{Z}_2$ compactification.

\section{Constraints on the gauge group of 8d and 9d theories}\label{sec:P}
We have seen how the ${\sf CP}$ actions that exists at a generic point on the Coulomb branch of 8d and 9d theories lead to constraints on the rank of the gauge group. As explained in Section \ref{sec:CP}, the only other piece of data involved in specifying an 8d/9d gauge theory is a choice of Lie algebra\footnote{Including massive states, we actually have a choice of a gauge group.} -- that is, one only needs to understand the possible symmetry enhancements at special points in the moduli space. We will now explore what the cobordism conjecture together with anomalies can say in this regard. 

 We will start by analyzing more general parity symmetries, like ${\sf P}$ that appears e.g. at the $SO(32)$ heterotic point. Any other parity symmetry in nine dimensions is equivalent to ${\sf CP}$ up to the action of an internal $\mathbb{Z}_2$ discrete gauge symmetry. ${\sf P}$ actions are is only relevant at the level of massive states or at points of enhanced symmetry, because on the generic point of the Coulomb branch the 9d supergravity theory does not have internal $\mathbb{Z}_2$ gauge symmetries other than those in the duality group.

We can get ${\sf P}$ from ${\sf CP}$ whenever we have internal charge conjugation symmetry ${\sf C}$ realized as a symmetry of the full theory. While in perturbative heterotic examples we can check explicitly whether a given ${\sf C}$ is a symmetry or not by just looking at the charge lattice, there is no recipe to do this in general. But in many cases ${\sf C}$ is in a sense actually part of the non-abelian gauge group. In these cases, it will automatically be a symmetry. 

To figure out when does this happen precisely, pick a Cartan subalgebra. ${\sf C}$ then acts on the weight lattice by sending each vector to minus itself. This is always an automorphism of the Lie algebra, and in some cases (in particular, every time the group only has real representations) it is a $\mathbb{Z}_2$ element of the Weyl group. The Weyl group $\mathcal{W}$ is not isomorphic in general to a subgroup of the gauge group, but rather it fits in a short exact sequence
\begin{equation}
\begin{tikzcd}1\arrow{r}&T\arrow{r}&N\arrow{r}&\mathcal{W}\arrow{r}&1\label{ses}\end{tikzcd}\end{equation}
where $T$ is the maximal torus associated to the Cartan subalgebra and $N$ its normalizer. When the sequence does not split, a lift of a generic Weyl group element to $G$ will not have the same order in general. We want a true $\mathbb{Z}_2$ gauge symmetry in order to keep the I-fold action of order two.  When part of the Weyl group, the question of whether ${\sf C}$ has a lift of order two was completely answered in \cite{2016arXiv160800510A}. To make a long story short, this is always the case for simple Lie groups for which all representations are real. 

Suppose we have such a simple Lie group $G$. Then we proceed as above, obtaining a pure ${\sf P}$ action. Let us work out the anomaly constraint for these.  At points in the moduli space with enhanced symmetry, a gauge group with a non-abelian factor $G$ for which charge conjugation is a symmetry will lead to vector multiplets and a $G$ gauge symmetry in six dimensions. The rest of the abelian factors contribute $1+8k-\text{rank}(G)$ hypermultipletswhere $k=0,1,2$ depending on whether the rank is $1,9$ or 17.  So the change in anomaly cancellation condition leads to 

\begin{equation}\text{dim}(G)+\text{rank}(G)\,\equiv\,0\,\text{mod}\,8.\label{mod8c}\end{equation}
We are thus making the prediction that for non-abelian factors $G$ which only admit real representations, \eq{mod8c} is a Swampland constraint. Let us check this against heterotic constructions. The following table lists the quantity \eq{mod8c} for the groups one can get from simple Lie algebras\footnote{For $\mathfrak{su}(k+1)$, many more cases are possible, e.g. $SU(k+1)/\mathbb{Z}_l$ makes sense whenever $l$ divides $k+1$.}:
\begin{table}[!htb]
\begin{equation*}
\begin{array}{c|c|c|c|c}
\text{Algebra}&\text{dim}(\mathfrak{g})+\text{rank}(\mathfrak{g})&\neq \text{0 mod 8?}&\text{Group}&\text{Real reps?}\\\hline
 \multirow{2}{*}{$A_{k}$}& \multirow{2}{*}{$k^2+3k$}&\multirow{2}{*}{$k\neq0,5\,\text{mod}\,8$}&SU(k+1)&\mxmark\\
&&&\cellcolor{swamp}PSU(k+1)&\cellcolor{swamp}\mcmark\\\hline
 \multirow{2}{*}{$B_{k}$}& \multirow{2}{*}{$2k(k+1)$}&\multirow{2}{*}{$k\equiv1,2 \,\text{mod}\,4$}&Spin(2k+1)&\mxmark\\
&&&\cellcolor{swamp}SO(2k+1)&\cellcolor{swamp}\mcmark\\\hline
 \multirow{2}{*}{$C_{k}$}& \multirow{2}{*}{$2k(k+1)$}&\multirow{2}{*}{$k\equiv1,2 \,\text{mod}\,4$}&Sp(k)&\mxmark\\
&&&\cellcolor{swamp}Sp(k)/\mathbb{Z}_2&\cellcolor{swamp}\mcmark\\\hline
 \multirow{3}{*}{$D_{k}$}& \multirow{3}{*}{$2k^2$}&\multirow{3}{*}{$k$ odd}&Spin(2k)&\mxmark\\
&&&\cellcolor{swamp}SO(2k)&\cellcolor{swamp}\mcmark\\
&&&\cellcolor{swamp}Spin(2k)/\mathbb{Z}_4&\cellcolor{swamp}\mcmark\\\hline
 \multirow{2}{*}{$E_{6}$}& \multirow{2}{*}{$84$}&\multirow{2}{*}{\cmark}&E_6&\mxmark\\
&&&\cellcolor{swamp}E_6/\mathbb{Z}_3&\cellcolor{swamp}\mcmark\\\hline
 \multirow{2}{*}{$E_{7}$}& \multirow{2}{*}{$140$}&\multirow{2}{*}{\cmark}&E_7&\mxmark\\
&&&\cellcolor{swamp}E_7/\mathbb{Z}_2&\cellcolor{swamp}\mcmark\\\hline
E_8&256&\mxmark&E_8&\mcmark\\
F_4&56&\mxmark&F_4&\mcmark\\
G_2&16&\mxmark&G_2&\mcmark\\
\end{array}
\end{equation*}
\caption{Allowed global forms of the gauge group in 8d and 9d. Any entry shaded in red is in the Swampland, once one takes into account the constraints on the second column.}
\label{latabla}
\end{table}

Anything in the above table in an entry shaded in red does not satisfy \eq{mod8c} and thus is in the Swampland. More concretely, the groups in the entries in red can never arise as a subgroup of the 9d gauge group (one must remember to take into account the constraints in the third column). Our constraints apply also in 8d via compactification to 5d. 

Thus, we get constraints on the allowed global forms of the gauge group. These are all satisfied in known heterotic examples; see \cite{Font:2020rsk} for an exhaustive list, including partial information on the global structure of the gauge group. For instance, there is a point on the 9d moduli space where the Lie algebra is $\mathfrak{so}(34)$. The above predicts the Lie group has to be the full $\text{Spin}(34)$, which is the case (the other options would not have spinors, which we have in heterotic).

It has to be noted that one can construct groups with only real representations which have reducible Lie algebras. For instance
\begin{equation} \frac{E_7\times SU(2)}{\mathbb{Z}_2}\end{equation}
only has real representations, since the fundamentals of both $SU(2)$ and $E_7$ are pseudorreal, and the quotient identifies them to produce a real representation. This group actually arises in heterotic moduli space by adjoint breaking \cite{Fraiman:2018ebo,Font:2020rsk}, so it better have $\text{dim}(G)+\text{rank}(G)$ equal to a multiple of eight. And indeed, rank is 8 and dimension is 3+133=136, for a total of 144 which is a multiple of 8. 

Suppose we go to one of the special points in moduli space where a ${\sf P}$ action is available, corresponding to charge conjugation of some subgroup $G$ of the full 9d gauge group, and perform the corresponding $T^3/\mathbb{Z}_2$ construction. We end up in a 6d $(1,0)$ theory with a non-abelian gauge group including $G$. As is well-known \cite{Kumar:2010ru,Monnier:2018nfs}, six-dimensional theories are very strongly constrained by anomaly cancellation. Together with the eightfold permutation symmetry of the I-fold, these can be used to further constraints in gauge groups, and in particular, we show that $\mathfrak{f}_4$ is inconsistent in nine dimensions. This is not new information \cite{Garcia-Etxebarria:2017crf}, but this result provides a nice consistency check of our approach. Details can be found in Appendix \ref{app:6danom}.

The above story can be generalized beyond pure parity actions. In general, assuming the existence of ${\sf CP}$ where ${\sf C}$ is some internal $\mathbb{Z}_2$ action that acts as charge conjugation when restricted to the Lie algebra (not necessarily part of the gauge group), we will be able to construct new ${\sf C'P}$ actions where ${\sf C'}={\sf C}\,  {\sf g}$ and ${\sf g}$ is a $\mathbb{Z}_2$ gauge transformation that commutes with ${\sf C}$. Thus, we must classify $\mathbb{Z}_2$ subgroups of Lie groups. Such a subgroup always lies in some Cartan torus, and it is uniquely identified by a choice of $r$  eigenvalues (where $r$ is the rank), all of which are either $+1$ or $-1$, modulo the action of the Weyl group on the Cartan. The techniques in \cite{Harvey:2017rko,2016arXiv160800510A} could be used to explore this more systematically. For instance, we haven't explored all the $\mathbb{Z}_2$ Weyl group elements which lift to $\mathbb{Z}_2$ gauge transformation. All the data to do this can be found in reference \cite{2016arXiv160800510A}. It is likely this leads to additional constraints.

\section{Comments on charge lattices}\label{sec:lattices}
In the previous Sections we have seen how a combination of the cobordism conjecture and anomaly cancellation can put some theories in the Swampland. To do this, however, we had to make assumptions -- notably, we assumed parity to be a good symmetry, and also charge conjugation in the eight-dimensional example. In this Section we will justify these assumptions from a more general perspective.

A theory of gravity  with a Minkowski vacuum and 16 supercharges in any number of dimensions has a very constrained structure. All a low-energy observer cares about is the physics everywhere in the moduli space, which is always locally a Narain coset. Usually, to specify a theory in Minkowski space we would need to specify the set of asymptotic states as well as the S-matrix. However, with so much supersymmetry, even the full massive spectrum of the theory is in some sense redundant. In what follows, we will assume that the low energy physics everywhere in moduli space is uniquely captured by the charge lattice. This is the case in  all known stringy examples, and the basic consequence of this assumption is that 
\begin{equation} \textbf{Symmetries of the lattice}=\textbf{Symmetries of the theory}.\end{equation}
The Narain charge lattice is the more familiar sixteen-supercharge analog of the situation with eight supercharges, where the general data encoding a supergravity theory  can be recast in terms of mixed Hodge structures (which has been proposed to be universal \cite{Grimm:2018ohb,Cecotti:2020rjq}) in that it captures the low-energy physics everywhere in moduli space.

The full data of the supergravity theory at any point in moduli space is fully specified by the charge lattice  $\Gamma_{10-d,k}$, of signature $(10-d,k)$ with $k\geq d-10$, and a plane $V_k$ in $\Gamma_{10-d,k}\otimes \mathbb{R}$ of signature $(0,k)$. When $V_k$ contains a  sublattice of  $\Gamma_{10-d,k}$, the corresponding gauge bosons are massless. See \cite{Mikhailov:1998si} for a discussion in the context of the CHL string, and \cite{Fraiman:2018ebo} for an exhaustive survey of heterotic compactifications to 9d in terms of the charge lattice. 

The duality group of the theory is then simply $O(\Gamma_{10-d,k})$, the group of indefinite orthogonal transformations that preserve the lattice. Similarly, ${\sf CP}$ is a symmetry of the low-energy lagrangian that does not act on the charge lattice, since the scalars that parametrize $V_k$ are left invariant. As such, it is a symmetry of the theory. We may combine it with internal symmetries of the lattice: for instance, if the scalars are tuned so that there is a non-abelian symmetry group, the Weyl group will be part of the subgroup of lattice symmetries that preserve $V_k$.

This perspective also allows us to study I-folds which are not immediately related to gauge symmetries of the underlying theory. For instance, on the rank 18 component of the moduli space of known 8d theories, there is a point of enhanced gauge symmetry
\begin{equation} E_8\times E_8 \times SU(3),\label{eeeee}\end{equation}
where there is also a $\mathbb{Z}_3$ R-symmetry, coming from unbroken automorphisms of the charge lattice. We can compactify this theory on $T^2/\mathbb{Z}_3$ combined with this R-symmetry to preserve supersymmetry in the bulk. There are 3 fixed points, and so we get a modulo 3 condition. We get one tensor and one hypermultiplet instead of three from the gravity multiplet, since the $T^2$ complex structure is fixed. The $E_8\times E_8$ vectors contribute 496 6d vectors, but for $SU(3)$ we must take into account that the R-symmetry acts as the Weyl group element on $SU(3)$ weight lattice given by $2\pi/3$ rotation. The Cartans are projected out, and out of the 2 groups of three roots in the orbits of this rotation, only the invariant combinations are projected in, producing 6d vectors. The 6d gravitational anomaly  condition becomes, taking into account the contribution of twisted sectors,
\begin{equation} 29+1-496-2\equiv\,273\,\text{mod}\,3,\end{equation}
so the I-fold compactification predicts \eq{eeeee} is allowed, as it should. We can now combine the R-symmetry action with a Weyl transformation of $SU(3)$, undoing the effect of the R-symmetry there. This has the effect of producing 8 vectors instead of 2; since 8-2=6, the constraint is still satisfied, as above.

Given this, suppose one had an 8d model with $G_2$ gauge symmetry. And suppose the charge lattice is such that it is possible to tune it to a point in moduli space with an unbroken $\mathbb{Z}_3$ symmetry, as we did before. Then, one can compactify on $T^2/\mathbb{Z}_3$, twisting by the R-symmetry as above, producing a 6d model free of mod 3 anomalies. The $G_2$ gauge fields would produce $14$ vectors in 6d, by a reasoning similar to the $SU(3)$ case above. But one could also twist by the combination of R-symmetry and the $\mathbb{Z}_3$ in the Weyl group of $G_2$, producing 4 6d vectors instead.  This results in a change in the mod 3 anomaly of 
\begin{equation}14-4=10\equiv 1,\,\text{mod}\,3.\end{equation}
 Thus, if one could argue the existence of a $\mathbb{Z}_3$ R-symmetry anywhere on the $G_2$ branch of moduli space, the $G_2$ theory would be put in the Swampland. 

We conclude this Section by noting a few properties that lattices of known string compactifications have:\begin{itemize}
\item In nine dimensions, the three known examples have even, self-dual charge lattices. If we found an underlying reason for self-duality, we could have string universality in nine dimensions. One possibility is modular invariance of the worldsheet of the fundamental string. Notice that an odd self-dual lattice is inconsistent, since the $F_4$ weight lattice can be embedded into the eight-dimensional odd self-dual lattice. We know $F_4$ is inconsistent, and its ``sickness'' extends throughout all of the moduli space. 
\item In eight dimensions, non self-dual lattices are allowed (see \cite{Mikhailov:1998si}), but the only known example has index two. This allows for symplectic Lie algebras, but not for $\mathfrak{g}_2$.

\item The pattern of allowed ranks of the gauge group in seven dimensions acquires an interesting meaning from the lattice point of view. The allowed values of the rank can be rewritten in the suggestive form
\begin{equation} r=3+l\cdot 2^k,\quad l=0,1,\quad k=1,2,3,4.\end{equation}
In all known cases the lattice of the known seven-dimensional theories can be decomposed as
\begin{equation}\Gamma_{3,3+k}=\Gamma_{3,3}\oplus \tilde{\Gamma},\end{equation}
where $\Gamma_{3,3}$ is a self-dual lattice of signature $3,3$ and $\tilde{\Gamma}$ is a purely Euclidean lattice of dimension 0 (no lattice), 2, 4, 8 or 16. Currently we do not have an explanation for these powers of two.  It is also possible that there are other stringy constructions which fill the gap and lead to all the odd rank cases up to the maximum rank allowed.
\end{itemize}
We suspect that perhaps all the constraints we have discussed in this paper, and many more, can be derived more systematically from an understanding of the properties of these lattices from first principles, which we hope to study further.

\section{Conclusions}\label{sec:conclus}
String theory is the only consistent quantum theory of gravity we know of so far. Since it is the only example, at first sight it might seem dangerous to extrapolate the features we find in string compactifications to general quantum gravity models  -- we might be looking under a lamppost!  The String Lamppost Principle (SLP), which is one of the main motivations of the Swampland program (sometimes also referred to as ``string universality''), is the idea that  the general features of quantum gravity are correctly captured in string compactifications, and that in fact \emph{every} consistent quantum gravity arises in the string landscape. The lamppost is bright enough to illuminate the whole street!

In this paper, we have investigated the SLP for high-dimensional ($d>7$) theories with 16 supercharges. On a generic point of the moduli space, the low-energy supergravity is completely specified by a single integer, the rank $r$ of the gauge group. We have shown that absence of cobordism global symmetries leads to a modulo 8 periodicity on the allowed values of $r$. This,  together with the upper bound in the rank argued in \cite{Kim:2019ths}, is enough to establish the SLP -- the only solutions to the above constraints are those realized by known string compactifications, as illustrated in Table \ref{tabla2}. We find it remarkable that a few general quantum gravity principles (absence of global symmetries and cobordism classes, and the distance conjecture) are enough to limit the general structure of the existing theories to those constructible in string theory for $d>7$. 

\begin{table}[!hbt]
\begin{center}
\begin{tabular}{c|cccccccccccccccccccc}
$9$&\cellcolor{swamp}0&\cellcolor{landscape}1&\cellcolor{swamp}2&\cellcolor{swamp}3&\cellcolor{swamp}4&\cellcolor{swamp}5&\cellcolor{swamp}6&\cellcolor{swamp}7&\cellcolor{swamp}8&\cellcolor{landscape}9&\cellcolor{swamp}10&\cellcolor{swamp}11&\cellcolor{swamp}12&\cellcolor{swamp}13&\cellcolor{swamp}14&\cellcolor{swamp}15&\cellcolor{swamp}16&\cellcolor{landscape}17&\cellcolor{swamp}&\cellcolor{swamp}\\\hline
$8$&\cellcolor{swamp}0&\cellcolor{swamp}1&\cellcolor{landscape}2&\cellcolor{swamp}3&\cellcolor{swamp}4&\cellcolor{swamp}5&\cellcolor{swamp}6&\cellcolor{swamp}7&\cellcolor{swamp}8&\cellcolor{swamp}9&\cellcolor{landscape}10&\cellcolor{swamp}11&\cellcolor{swamp}12&\cellcolor{swamp}13&\cellcolor{swamp}14&\cellcolor{swamp}15&\cellcolor{swamp}16&\cellcolor{swamp}17&\cellcolor{landscape}18&\cellcolor{swamp}\\\hline
$7$&\cellcolor{swamp}0&\cellcolor{undecided}1&\cellcolor{swamp}2&\cellcolor{landscape}3&\cellcolor{swamp}4&\cellcolor{landscape}5&\cellcolor{swamp}6&\cellcolor{landscape}7&\cellcolor{swamp}8&\cellcolor{undecided}9&\cellcolor{swamp}10&\cellcolor{landscape}11&\cellcolor{swamp}12&\cellcolor{undecided}13&\cellcolor{swamp}14&\cellcolor{undecided}15&\cellcolor{swamp}16&\cellcolor{undecided}17&\cellcolor{swamp}18&\cellcolor{landscape}19\\
\end{tabular}

\end{center}
\caption{Improved version of Table \ref{tabla1}, including the constraints in this paper. Each row corresponds to a dimension above six, and the numbers in each cell indicate the possible values of the rank. Red means the corresponding theory in the Swamp, green means there is a known stringy construction, and yellow means not ruled out, but there is no known stringy embedding. Notice that there is no yellow left in 8d and 9d, which means the SLP has been fully realized in these dimensions.}
\label{tabla2}

\end{table}

An important ingredient in our argument was the fact that 8d and 9d supergravity makes sense on non-orientable manifolds of the $\text{Pin}^-$ type. This leads to discrete global symmetries associated to non-orientable cobordisms. The cobordism conjecture \cite{McNamara:2019rup} implies the existence of localized defects that mediate breaking of this global symmetry, and which we have dubbed \emph{I-folds} (for ``inversion-folds''). They are orbifolds or orientifolds in ordinary string compactifications, but the point is that they exist in general. Anomaly cancellation in I-fold backgrounds then leads to the constraints mentioned above.

We have also used I-fold compactifications to produce nontrivial constraints of the global structure of the 8d and 9d gauge groups. While strictly speaking this is not part of the supergravity data and cannot be detected via low-energy experiments, it is an important piece of data of the full theory. Our constraints are consistent with everything we know in string compactifications\footnote{As we were completing this paper, we became aware of a work which leads to restrictions on allowed gauge groups based on 1-form anomalies and which has some overlap with our results in Section \ref{sec:P} \cite{Cvetic:2020kuw}.}. In particular, they are consistent with the results of a systematic exploration of all the possible maximal symmetry enhancements for heterotic on $T^2$ in \cite{Font:2020rsk}. In the eight-dimensional case, there could be a top. GS mechanism, as e.g. in \cite{Garcia-Etxebarria:2017crf} (or \cite{Wang:2020xyo,Wang:2020gqr}). In this case we have a modulo 4 constraint, which is robust.  However in quantum gravity such a tQFT can only arise as a low-energy effective description of anomalous UV degrees of freedom. There is a Swampland of such effective tQFT's, which deserves systematic exploration. It could also be that supersymmetry somehow constrains the tQFT coupling. 

We expect that the basic idea in this note, that we could summarize  schematically as
\begin{center}Cobordisms\,+\,Anomalies\,=\,Swampland,\end{center}
has applications in many other contexts, such as the Swampland of 6d theories \cite{Kim:2019vuc,Taylor:2018khc}, or even the Standard Model. We hope to explore these in future work. There is, however, several interesting questions left in $d\geq 7$ and sixteen supercharges, some of which we briefly comment on:
\begin{itemize}
\item The parity symmetry we have used is a good symmetry of the low-energy supergravity, and can only be broken by the massive states. With sixteen supercharges, the physics of these massive states (and indeed, all the theory) is completely specified by an integer charge lattice. Symmetries of the lattice are symmetries of the theory, explaining our parity symmetry and more. Further constraining the properties of the possible charge lattice would be a way to narrow down even more the constraints on compactifications with 16 supercharges. It may also be that  the 8d and 9d parity symmetry is a consequence of supersymmetry alone, though we do not have a derivation. 

\item Relatedly, we have not been systematic in our exploration of I-folds. At particular points in moduli space we could have additional symmetries, we could repeat the construction in Section \ref{sec:P} but quotienting by other elements of the Weyl group, etc. These could very well lead to new constraints. 

\item In seven dimensions, our constraints are not strong enough to argue for SLP, since we only get a modulo 2 periodicity in the rank, while string compactifications allow for rank $3+k\cdot 2^n$, where $k=0,1$ and $n=0,1,2,3,4$.   It should be checked whether other stringy constructions fill the gap and lead to all ranks or whether this power of two pattern is a necessity;  if so perhaps it is related to some property of the charge lattices that can arise in seven dimensions. 

\item The only other piece of data that one needs in specifying a $d>6$ supergravity is the possible non-abelian enhancements in moduli space. The most interesting open question is whether $\mathfrak{g}_2$ and  $\mathfrak{sp}(n)$ of maximal rank can exist in 8 and 9 dimensions. We do not have a string construction of these, but the I-folds that we considered here do not rule them out either. For $\mathfrak{g}_2$ in particular we have explored ordinary gauge anomalies in a variety of compactification manifolds, as well as I-fold compactifications to lower dimensions, and we have not found an inconsistency so far.  We view this as a particularly interesting left over question in the high-dimensional Swampland program.
\end{itemize}

\subsubsection*{Acknowledgements}

We thank Markus Dierigl, Jacob McNamara, Kantaro Ohmori, Carlos Shabazi, Irene Valenzuela, Juven Wang, and Gianluca Zoccarato for valuable discussions and comments on the manuscript, and Pieter-Jan De Smet for pointing out several typos. We are especially grateful to I\~{n}aki Garc\'{i}a-Extebarria for many useful discussions and comments, as well as collaboration at an early stage of this work. 

 We gratefully acknowledge the hospitality of UC Santa Barbara KITP during the workshop ``The String Swampland and Quantum Gravity Constraints on Effective Theories''.  This research was supported in part by a grant from the Simons Foundation (602883, CV).   The research of CV was in addition supported by the NSF grant PHY-1719924.

\appendix

\section{\texorpdfstring{$\text{Pin}^-$}{Pin-} fermions in nine and six dimensions}\label{app:dimred}
In this Appendix we work out the parity transformation properties of the fermions in the 9d $\mathcal{N}=1$ multiplets and work out their dimensional reduction on the $T^3$ I-fold described in the main text.

\subsection{Parity action on the 9d multiplets}
The group $\text{Pin}^-(8,1)$ has two real spinor representations, distinguished by whether the product of reflecting all space and time coordinates is $+1$ or $-1$. We will denote the corresponding real spinor representations as  $\mathbf{16}_\pm$. This extends to any spinorial representation of odd dimension; for instance, the gravitino can be described as a Rarita-Schwinger field (a reducible vector-spinor $\psi_{\mu a}$, where $a$ are Clifford algebra indices) subject to the constraint $\Gamma_{ab}^\mu \psi_\mu^a=0$. There are two such real representations, which we denote as $\mathbf{128}_\pm$.

Without loss of generality, we can fix the supercharge to live in the $\mathbf{16}_-$ representation. The scalar in the gravity multiplet is {\sf CP}-even, which then fixes the gravitino and dilatino to live in the $\mathbf{128}_+$. There are actually two different $\mathcal{N}=1$ vector multiplets, distinguished by their transformation properties under parity. If the scalar is {\sf CP}-even, then the gaugino lives in the $\mathbf{16}_+$ and the vector is odd under ${\sf CP}$ (transforms as an axial vector under parity). If on the other hand the scalar is ${\sf CP}$-odd, the gaugino lives in the $\mathbf{16}_-$ and the vector is ${\sf CP}$-even (transforms as a regular vector under parity). Both cases are relevant in known examples of $\mathcal{N}=1$ 9d theories, because a $\mathbf{16}_+$ cannot get a $\text{Pin}^-$ invariant mass by itself, but a $\mathbf{16}_+$ and $\mathbf{16}_-$ together can. Thus, pairs of vector multiplets of opposite chiralities can actually get a mass; in fact, the vector multiplets describing the $W^\pm$ bosons at any point of enhanced symmetry (such as the $E_8\times E_8$ string on a circle) decompose precisely in such a pair. We have summarized the field content of each multiplet and their parity transformations in Table \ref{t2}.

\begin{table}[!hbt]
\begin{center}

\begin{tabular}{c|c|c}
Multiplet&Field&Parity\\\hline
\multirow{5}{*}{Gravity}&Graviton&Even\\
&Gravitino&Even\\
&2-form&Even\\
&1-form&Odd\\
&Dilatino&Even\\
&Scalar&Even\\\hline
\multirow{3}{*}{$\text{Vector}^+$}&1-form&Odd\\
&Gaugino&Even\\
&Scalar&Even\\\hline
\multirow{3}{*}{$\text{Vector}^-$}&1-form&Even\\
&Gaugino&Odd\\
&Scalar&Odd\\
 \end{tabular}
\end{center}

\caption{Parity transformation of all the fields in the gravity and the two possible 9d $\mathcal{N}=1$ vector multiplets. ``Odd'' or ``even'' refers to whether the field picks an additional minus sign under a simultaneous reflection of all spatial and time coordinates.}
\label{t1}
\end{table}

As explained in Section \ref{sub:par}, we use the action of an internal duality symmetry, if necessary, to take the scalars in every vector to be odd. This corresponds to what we called ${\sf CP}$ symmetry. With an action of ${\sf P}$, some scalars are even and some are odd instead.

\subsection{Dimensional reduction of fermions on toroidal I-fold} 
Under the decomposition $\text{Pin}^-(8,1)\rightarrow \text{Spin}(5,1)\times \text{Pin}^-(3)$, we have
\begin{equation} \mathbf{16}_\pm\,\rightarrow [\mathbf{4}\otimes \mathbf{2}_\pm]^\mathcal{R}\oplus [\mathbf{4}'\otimes \mathbf{2}_{\pm}]^\mathcal{R},\label{flafy}\end{equation} 
where $\mathbf{4}$ and $\mathbf{4}'$ are the two (self-conjugate) conjugate Weyl representations of opposite chiralities of $\text{Spin}(5,1)$ and  $\mathbf{2}_+$ and $\mathbf{2}_-$ are the two inequivalent irreducible representations of $\text{Pin}^-(3)$ \cite{Witten:2015aba}. The notation $[\ldots]^{\mathcal{R}}$ means we have imposed a reality condition. In our case, this is possible because both the $\mathbf{2}_\pm$ and the $\mathbf{4},\mathbf{4}'$ are pseudorreal.  Similarly, the gravitino decomposes as
\begin{equation} \mathbf{128}_+\,\rightarrow [\mathbf{4}\otimes \mathbf{2}_+]^\mathcal{R}\oplus [\mathbf{4}'\otimes \mathbf{2}_-]^\mathcal{R}]\oplus [\mathbf{4}\otimes \mathbf{4}_+]^\mathcal{R}\oplus [\mathbf{4}'\otimes \mathbf{4}_-]^\mathcal{R}\oplus [\textbf{20}\otimes \mathbf{2}_+]^\mathcal{R} \oplus [\overline{\textbf{20}}\otimes \mathbf{2}_-]^\mathcal{R},\end{equation}

Imposing the orientifold projection means that only fields with a $\mathbf{2}_-$ or $\mathbf{4}_-$ of $\text{Pin}^-(3)$ survive the quotient. More concretely, the equation for the spinors in the compact space is
\begin{equation}\psi(-\vec{x})=-\Gamma^1\Gamma^2\Gamma^3\psi(\vec{x}),\label{we4we}\end{equation}
where $\vec{x}$ is a coordinate in the parent $T^3$ and $\pm$ according to whether we consider fermions on the $\mathbf{2}_\pm$ representation. In the 3d Dirac representation, the product $\Gamma^1\Gamma^2\Gamma^3$ is a multiple of the identity, $\pm\mathbf{I}$, which we take to be $\mathbf{I}$ in the $\mathbf{2}_+$ and $-1$ in the $\mathbf{2}_-$. Therefore, \eq{we4we} has either zero or two solutions. The same argument with the $\mathbf{4}$ leads to either zero or four solutions.
A representation like $[\mathbf{4}\otimes \mathbf{2}_+]^\mathcal{R}$ transforms a single complex $\mathbf{4}$ under the $\text{Spin}(5,1)$ subgroup.
Thus we get 
\begin{equation}\mathbf{128}_+\rightarrow \textbf{20} + 3\,(\mathbf{4}),\quad 
\mathbf{16}_+\rightarrow \mathbf{4},\quad \mathbf{16}_-\rightarrow \mathbf{4}'. \end{equation}
This is the fermionic content of the multiplets obtained in the main text from the reduction of the bosons.

 \section{Real fermions and Pin structures}\label{app:ferms}
Here we discuss some properties of Euclidean fermions which are relevant to the main results in the text, as well as the general relation between Lorentzian and Euclidean fermions and representations of the Pin group. An excellent reference is \cite{2001RvMaP..13..953B}.

Suppose one has a Lorentzian theory with fermions, and one wishes to extend the symmetry group to include reflections or time reversal. Parity is implemented on the spinor bundle as multiplication by a $\Gamma$ matrix times a phase \cite{Witten:2015aba}, so for instance
\begin{equation}{\sf P}^1:\quad \psi(x_1)\rightarrow e^{i\varphi}\, \Gamma^1 \psi(-x_1)\label{pph}\end{equation}
for a reflection along a coordinate $x_1$. Depending on whether the phase $e^{i\varphi}$ is real or imaginary, we get two different covers of the orthogonal group, the $\text{Pin}^\pm$ groups. In $\text{Pin}^+$, reflections square to the identity; in $\text{Pin}^-$, they square to $(-1)^F$. 

So given a particular theory, with some parity symmetry in the bosonic sector, can we extend it to $\text{Pin}^+$, $\text{Pin}^-$, or both? A simple example of a restriction comes from representation theory; as discussed in \cite{2001RvMaP..13..953B}, it can happen that $\text{Pin}^+$ has no real representations but $\text{Pin}^-$ does, or vice-versa. So whenever the theory we want to put in non-orientable manifolds contains fermions in real (Majorana) representations, only those $\text{Pin}$ covers which admit real representations will do. The results from \cite{2001RvMaP..13..953B} in Lorentzian signature are given in Table \ref{t2}, where we list which $\text{Pin}$ structures are compatible with a Majorana representation. For instance, in 11 dimensions the Majorana representation of the M-theory supercharge is only compatible with a $\text{Pin}^+$ structure\footnote{On non-orientable manifolds, this is only possible by having a bundle with external Clifford multiplication \cite{Lazaroiu:2016nbq}, that is, there are two complementary $\text{Pin}^+$ structures on the manifold and the Dirac operator maps one to the other.}, and in 9 and 8 dimensions (the cases of interest in this paper) the real supercharge is only compatible with a $\text{Pin}^-$ structure. 

\begin{table}[!hbt]
\begin{center}

\begin{tabular}{c|cccccccccccc}
$d$&$1$&$2$&$3$&$4$&$5$&$6$&$7$&$8$&$9$&$10$&$11$&$12$\\\hline
$\text{Pin}^+$&\xmark&\cmark&\cmark&\cmark&--&--&--&\xmark&\xmark&\cmark&\cmark&\cmark\\\hline
$\text{Pin}^-$&\cmark&\cmark&\xmark&\xmark&--&--&--&\cmark&\cmark&\cmark&\xmark&\xmark
 \end{tabular}
\end{center}

\caption{Possible Pin structures compatible with a Majorana condition in Lorentzian signature. A dashed line indicates that no Majorana condition is possible. The whole table is periodic modulo 8 in $d$, and works for any signature if one replaces $d$ by $s-t+1$, where $s$ is the number of dimensions with $+$ signature and $t$ the number of dimensions with $-$ signature. Note however that Majorana conditions on spinors are not directly relevant for reflection-positive Euclidean theories.}
\label{t2}
\end{table}

Upon Wick rotation of the time coordinate, time reversal becomes just another reflection. Hence, Table \ref{t1} controls the $\text{Pin}$ structure of the Euclidean version of the theory as well. This leads to an apparent puzzle: Euclidean spinor representations in a given dimension have very different reality properties from Lorentzian ones; for instance, in 11 dimensions the smallest representation of the Euclidean Clifford algebra is complex 32-dimensional instead of real 32-dimensional. The trick is that in Euclidean signature  the supercharges are no longer required to be hermitian, but reflection positive. More concretely,
\begin{equation} Q^\dagger_a= ({\sf P}^0)_{a}^b Q_b= e^{i\varphi} (\Gamma^0)_a^b Q_b.\label{rpfer}\end{equation}
By construction, this leads to the same reality conditions and Pin covers as in Lorentzian signature. For instance, in 9 Euclidean dimensions, the $\Gamma$ matrices can be taken to be real and symmetric. Then, acting with a reflection twice and requiring that \eq{rpfer} stays invariant leads to $e^{2i\varphi}=-1$, which corresponds to a $\text{Pin}^-$ structure. 

At the level of a fermionic path integral, the counting of fermionic degrees of freedom works in the same way. For a given Dirac operator, there are \emph{two} a priori independent fermionic variables in the path integral, $\psi$ and $\tilde{\psi}$,\begin{equation} Z=\int \mathcal{D}\psi\mathcal{D}\tilde{\psi} \exp\left(-\int d^dx \tilde{\psi}^T\slashed{D}\psi\right),\label{fpi}\end{equation}
 In Lorentzian signature it is customary to write $\tilde{\psi}$ as $\psi^*$, but this is a misnomer as there is no natural notion of ``conjugating a fermionic variable''. So for instance, in 11 dimensions, in the expression \eq{fpi} there are actually 64 fermionic variables we path integral over. In some cases however, it is possible to impose conditions relating $\tilde{\psi}$ and $\psi$, thus reducing the number of variables.  For instance, for an 11d Dirac spinor\footnote{The whole discussion actually applies to the M-theory gravitino, since the extra vector index does not play a role.} one can impose a Majorana condition in Lorentzian signature,
 \begin{equation}\psi= \mathcal{B}_+\psi,\quad \tilde{\psi}= \mathcal{B}_+\tilde{\psi}\label{mcond}\end{equation}
 where the symmetric matrix $\mathcal{B}_-$ satisfies $\mathcal{B}_-\mathcal{B}^*=1$\footnote{A great reference is \cite{Ortin:2015hya}, Appendix D, which contains all the relevant information on spinors and reality conditions of any signature and dimension.}. This cuts down the number of fermionic variables to the expected 32\footnote{The ``twisted'' Majorana condition $\tilde{\psi}=\mathcal{B}_+\psi$ would cause the Dirac lagrangian to vanish identically.}. In Euclidean signature, there is an antisymmetric matrix that satisfies $\mathcal{B}_-\mathcal{B}^*=-1$ instead, so the Majorana condition cannot be imposed. However, two wrongs can make a right \cite{polchinski1998string}; we can introduce an internal map that sends
\begin{equation}\Omega:\, \chi=\left(\begin{array}{c} \psi\\\tilde{\psi}\end{array}\right)\quad\rightarrow\quad \left(\begin{array}{cc} 0&-\mathcal{B}^*_\pm\\\mathcal{B}_\pm&0\end{array}\right)\left(\begin{array}{c} \psi\\\tilde{\psi}\end{array}\right),\label{omegadef}\end{equation}
which as on the (reducible) representation of the Clifford algebra which is the direct sum of $\psi$ and $\tilde{\psi}$. And on this representation, we can impose the condition
\begin{equation}\chi=\Omega \chi.\label{SMcond}\end{equation}
This condition also cuts down the number of fermionic degrees of freedom by half, so again we obtain the right counting. 

Finally, we quickly comment on the existence of additional internal charge conjugation symmetries for irreducible Majorana representations. These can be obtained from the general discussion for Clifford algebras and can be found on any standard reference \cite{polchinski1998string,Ortin:2015hya}. These results imply in particular that there is no possible extension of the $\text{Pin}^-$ group in $\mathcal{N}=1$ theories in nine dimensions  since  the supercharge does not admit any additional internal symmetries. In eight dimensions the only possibility for a single Majorana spinor is the charge conjugation symmetry ${\sf \tilde{C}}$ of Section \ref{sec:par8d0}.

\section{Supersymmetry and parity anomaly in one dimension}\label{app:parity1d}
In Section \ref{sec:tach}, we used supersymmetry to argue that the charge of the I5-fold is integer-quantized. The argument is basically that due to the large amount of supersymmetry, the contribution to the parity anomaly of a whole multiplet vanishes. Here we offer a more general argument, without relying on free fields.

In general, anomalies are only sensitive to gapless degrees of freedom. For a 1d quantum system as the worldvolume of the electric brane discussed in Section \ref{sec:tach}, this means that it only depends on the (finite-dimensional) Hilbert space of groundstates. This space of groundstates must furnish a representation of the symmetry group, including symmetries such as time reversal and parity, if they exist. 

Part of the symmetry group could be anomalous, which for a one-dimensional quantum system like the one we have means that the representation wouldn't be linear, but projective. Studying anomalies one-dimensional systems with a $\text{Pin}^-$ symmetry means classifying projective representations of an antiunitary time-reversal symmetry ${\sf T}$ that squares to the identity. The classification was carried out in  \cite{2011PhRvB..83g5103F} and there are eight cases, corresponding to the 8 elements of the bordism group $\Omega_2^{\text{Pin}^-}=\mathbb{Z}_8$. Only one of them is not projective, corresponding to the anomaly-free case. Our task is to figure out which of these projective representations can be realized when coupled to eight supercharges. For us, the supercharges are there from the 9d perspective; see \cite{1dSUSY} for an emergent realizations of supersymmetry direclty in 1d.

The 9d supersymmetry algebra contains real (hermitean) supercharges $Q_\alpha$, that satisfy \cite{AbouZeid:1999fv}
\begin{equation}\{Q_\alpha,Q_\beta\}= i\slashed{P}\Gamma^0+ Z\,\Gamma^0,\end{equation}
where $Z$ is a real central charge. For a massive particle at rest, $P_\mu=(m,0,0,\ldots)$. The supersymmetry algebra takes the form 
\begin{equation}\{Q_\alpha,Q_\beta\}= m\mathbf{I}+Z\Gamma^0\,\end{equation}
where $(\Gamma^0)^2=1$. It is now convenient to decompose this equation into irreps of the $\text{Spin}(8)$ little group of the BPS particle. $\Gamma^0$ has eigenvalues $\pm1$, so the above equation becomes the sum of two $8\times8$ blocks, with eigenvalues $Z\pm m$. If the particle is BPS, one of these eigenvalues vanishes, and corresponds to the eight unbroken supercharges. In any case, the other $8\times8$ block always describes eight unbroken supercharges $Q_a$, where $a=1,\ldots 8$, are hermitean $Q_a=Q_a^\dagger$ and satisfy the commutation relations
\begin{equation}\{ Q_a,Q_b\}= Z\,\delta_{ab}.\label{clll}\end{equation}
The tensor $\delta_{ab}$ is the  symmetric invariant tensor of the Majorana-Weyl spinor representation of the little group $\text{Spin}(8)$ of a massive particle. 

Unlike in higher dimensions, the ``broken'' supercharges are represented linearly in the worldvolume  theory, since there is no spontaneous symmetry breaking in quantum mechanics. 

Equation \eq{clll} is the 8d Euclidean Clifford algebra, and so the groundstate space of the worldvolume theory furnishes a (nontrivial) real (because the supercharges are hermitean) representation of the 8d Clifford algebra. There is only one such nontrivial irrep \cite{Ortin:2015hya}, of real dimension 8, so the groundstate space decomposes as a direct sum of a number of copies of it, and a priori we might think there are possibly some singlets too.  However, there can be no trivial reps. of the Clifford algebra in the groundstate space, since these would correspond to states preserving all 16 supercharges. The only such state is the vacuum, with $Z=0$. 

Let us consider a single irreducible block. In this block, the supercharges $Q_a$ transform under parity and time reversal in the exact same way as the groundstate space of the free, $\mathcal{N}=8$ 1d multiplet, which has no parity anomaly. In more abstract terms, an antilinear operator {\sf T} acting on the supercharges as 
\begin{equation} {\sf T} Q_a {\sf T}^{-1} =Q_a\label{parcc}\end{equation}
must necessarily satisfy ${\sf T^2}=+1$, since otherwise we could combine it with charge conjugation to construct a linear parity symmetry of the Clifford algebra that would square to $+1$ \cite{Witten:2015aba}. As discussed in Appendix \ref{app:ferms}, this is not possible; only $\text{Pin}^-$ is possible. Time reversal is thus represented linearly. Fermion number is also represented linearly and the representation of the algebra is even, so the representation is anomaly free according to the criteria in \cite{2011PhRvB..83g5103F}.  Thus, the general groundstate cannot have a parity anomaly either. 

This line of reasoning allows one to rule out would-be worldvolume theories which a priori seem completely nice, but which could have a parity anomaly. For instance, one could have the eight worldvolume goldstinos discussed in the main text (which have no parity anomaly) coupled to the edge mode of a Haldane spin chain \cite{2011PhRvB..83g5103F}, a bosonic phase which realizes the generator of $\Omega_{2}^{\text{Unoriented}}=\mathbb{Z}_2$. This has the same $\text{Pin}^-$ anomaly as four Majorana fermions, and so the resulting theory would contribute a sign to the Dirac phase of the electric brane, shifting the quantization condition of the I5 to half-integers. However, as we have seen, this coupling is incompatible with supersymmetry realized via hermitean supercharges. The resulting ${\sf T}$ operator would necessarily map different Clifford blocks to each other, contradicting \eq{parcc}.

There is a similar loophole related to M2 branes, which is not so easily solved since the relevant system is 3-dimensional instead of 1-dimensional. It was argued in the classical paper \cite{Witten:1996md} (see also \cite{Witten:2016cio}) that the worldvolume theory of the M2 has a parity anomaly, leading to a fractional charge for the MO5-plane. But what if the M2 had an additional worldvolume coupling with anomaly given by e.g. eight times that of the generator of $\Omega_{4}^{\text{Pin}^+}=\mathbb{Z}_{16}$? This would provide an additional sign contribution to the parity anomaly, leading to MO5-planes with integer-quantized charge. This anomaly theory can be realized by a Chern-Simons tQFT, so in particular it is clear that we can realize it without changing the massless spectrum of the M2 brane. In fact, any $\Omega_{4}^{\text{Pin}^+}$ anomaly theory can be realized via Chern-Simons theories \cite{Cordova:2017vab,Wang:2020xyo}, so it would seem we can shift the MO5 charge quantization to any multiple of $1/16$. Our suspicion is that the $\mathcal{N}=8$ supersymmetry in the M2 worldvolume prevents this, possibly by placing constraints on the quantum numbers of topological operators in the theory (anyons). The anyon spectrum is directly related to the parity anomaly of the theory, via an ``anomaly indicator'' \cite{Tachikawa:2016cha,Wang:2016qkb,Tachikawa:2016nmo}. Also, we would expect operators in the bulk theory to correspond to dynamical objects in 11 dimensions, and these should fall in supersymmetry multiplets. 

\section{6d anomaly cancellation}\label{app:6danom}
Suppose we go to one of the special points in moduli space where a ${\sf P}$ action is available, corresponding to charge conjugation of some subgroup $G$ of the full 9d gauge group, and perform the corresponding $T^3/\mathbb{Z}_2$ construction. We end up in a 6d $(1,0)$ theory with a non-abelian gauge group including $G$. As is well-known \cite{Kumar:2010ru,Monnier:2018nfs}, six-dimensional theories are very strongly constrained by anomaly cancellation. Here we see how the requirement of anomaly cancellation, together with additional structure of the 6d theory that comes from the eightfold permutation symmetry of $T^3/\mathbb{Z}_2$, can put constraints on the non-abelian factors $G$ that can arise in 9 dimensions. 

We will only focus on gravitational and mixed $G$-gravitational anomalies of the 6d theory. There might be further constraints involving e.g. mixed anomalies with other non-abelian factors or with abelian ones. As before, we will use the technology  and language of anomaly cancellation in $(1,0)$ theories, even if the I-fold might not always be supersymmetric. The reason is again that the only additional ingredient that supersymmetry breaking brings is that there can be an effective number of hypermultipletsin each representation of the gauge group, which could possibly be negative in charge. Since all the constraints we look at are modulo some integer, this won't affect our results. 

In a 6d theory with $T$ tensor multiplets, anomalies are encoded by two vectors $\vec{a}$ and $\vec{b}$ whose coefficients correspond to the Green-Schwarz couplings in the 6d Lagrangian. $\vec{a}$ is related to cancellation of gravitational anomalies while $\vec{b}$ relates to cancellation of pure gauge and mixed anomalies; in general, we have one such vector $\vec{b}$ for each non-abelian factor of the gauge group, but it is consistent to focus on the set of constraints one gets for a single vector. 

In a general theory, on top of the vector multiplet in the adjoint of $G$, we will have a number of hypers, in generic representations $\mathbf{R}$ of the gauge group. The anomaly vectors $\vec{a},\vec{b}$ must belong to a self-dual lattice equipped with an inner product of signature $(1,T)$. This integrality condition is related to the Dirac pairing of the self-dual strings that couple to the tensors \cite{Kumar:2010ru,Monnier:2018nfs}. Anomaly cancellation constrains the inner products of the vectors as  \cite{Kumar:2010ru,Monnier:2018nfs}
\begin{align} \vec{a}\cdot\vec{a}&=9-T, \nonumber\\ \vec{a}\cdot\vec{b}&=\frac{\lambda}{6}\left(A_{\text{adj.}}-\sum_{\mathbf{R}} n_{\mathbf{R}}A_{\mathbf{R}}\right),&\vec{b}\cdot\vec{b}=-\frac{\lambda^2}{3}\left(C_{\text{adj.}}-\sum_{\mathbf{R}} n_{\mathbf{R}}C_{\mathbf{R}}\right).\label{anocanc}\end{align}
Here, $n_{\mathbf{R}}$ is the number of hypermultipletsin representation ${\mathbf{R}}$. It might seem that to analyze anomalies in general, one would need to discuss all (infinitely many) possible representations of the gauge group, but fortunately this is not necessary.  A Lie algebra $\mathfrak{g}$ of rank $k$ has $k$ fundamental representations $\mathbf{r}_i$. Any other representation arises in the tensor product of some of the fundamental representations. This implies that
\begin{equation} \left(\begin{array}{c} A_{\mathbf{R}}\\C_{\mathbf{R}}\end{array}\right)=\sum_{\mathbf{r}_i} s^{\mathbf{R}}_i  \left(\begin{array}{c} A_{\mathbf{r}_i}\\C_{\mathbf{r}_i}\end{array}\right),\quad s_i\in\mathbb{Z}\end{equation}
so that the anomaly vector for any linear combination of representations can be replaced by effective numbers in the fundamental representations,
\begin{equation}  \sum_{\mathbf{R}} n_{\mathbf{R}}\left(\begin{array}{c} A_{\mathbf{R}}\\C_{\mathbf{R}}\end{array}\right)\quad\rightarrow \quad  \sum_{\mathbf{r}_i}  n'_{\mathbf{r}_i} \left(\begin{array}{c} A_{\mathbf{r}_i}\\C_{\mathbf{r}_i}\end{array}\right),\end{equation}
where $n'_{\mathbf{r}_i}$ is some effective integer number of hypermultipletsin representation $\mathbf{r}_i$. 

The analysis so far has been completely general -- let us now specialize to the $T^3/\mathbb{Z}_2$ case. New restrictions arise due to the eightfold permutation symmetry of the orbifold fixed loci. The additional hypermultiplets and tensors are all localized in the orbifold fixed loci, and since each contributes the same fields, the total contribution must be a multiple of eight:
\begin{equation} n'_{\mathbf{r}_i}=8 k_i,\quad k_i\in\mathbb{Z}.\end{equation}
On the other hand the string charge lattice also has additional structure due to the permutation symmetry. One gets one self-dual tensor from the 9-dimensional $B$-field, and $8t$ tensors from the orbifold fixed loci, so $T=1+8t$.  The string charge lattice is therefore a self-dual lattice $\Gamma_{1,1+8t}$ of signature $(1,1+8t)$. We will  denote a generic vector in the charge lattice as
\begin{equation} (\vec{v}, \vec{m}_1, \vec{m}_2\ldots)\label{vch}\end{equation}
where $\vec{m}_i$ is a $t$-dimensional vector labelling the charges associated to self-dual tensors in the $i$-th fixed locus, and $\vec{v}$ is a two-dimensional vector labeling electric and magnetic charges under the supergravity $B$-field. The action of the symmetric group $\mathcal{S}_8$ on the eight orbifold fixed points acts on the charge lattice by permuting the corresponding components $\vec{m}_i$ in \eq{vch}, and leaving the first two invariant.

Furthermore, since each $\mathbb{R}^3/\mathbb{Z}_2$ singularity is independent of each other, we will take the charge lattice to be generated by vectors of the form
\begin{equation} (v_1,v_2,0,\ldots \vec{m}_i,\ldots,0)\quad\text{and}\quad (v_1,v_2,0,\ldots,0).\end{equation}
The first one corresponds to strings charged under the self-dual tensors of the $i$-th locus, and possibly the supergravity fields as well. The second are the supergravity strings. The key assumption we are making is that there are no additional strings charged under several of the fixed loci at the same time (other than multi-string states). 

 This $\mathcal{S}_8$ action must also respect the inner product and severely constrains it. Vectors of the form
\begin{equation} (0,0,\ldots \vec{m}_i,\ldots)\quad\text{and}\quad (0,0,\ldots \vec{m}_j,\ldots)\end{equation}
for $i\neq j$ must be orthogonal, since we could act with a permutation leaving fixed the $i$-th vector and moving around the other. Similarly, an inner product 
\begin{equation} (v_1,v_2,0,\ldots) \cdot (0,0,\ldots \vec{m}_j,\ldots)\end{equation}
must be independent of $j$. Finally, we also know the upper left $2\times 2$ block of the inner product, since this describes the restriction to the supergravity sector, where the Dirac pairing is known exactly from dimensional reduction from 9d: it is simply 
\begin{equation} \left(\begin{array}{cc}0&1\\1&0\end{array}\right),\end{equation}
the inner product of the even and self-dual Lorentzian lattice $\Gamma_{1,1}$. Taking everything into account, the matrix of the inner product in the above basis must take the form 
\begin{equation}\left(\begin{array}{ccccc}0&1&\vec{m}_1& \dots& \vec{y_1}\\ 1&0&\vec{y}_2& \dots& \vec{y_2}\\\vec{y_1}&\vec{y}_2& \mathbf{G} & \dots &0\\\vdots&\vdots&\vdots& \ddots & \vdots\\\vec{y_1}&\vec{y_2}&0& \dots & \mathbf{G}\end{array}\right)\label{sese}\end{equation}
where the vectors $\vec{y}_1,\vec{y}_2$ have integer inner products with all $\vec{m}$ and the symmetric $t\times t$ matrix $\mathbf{G}$ specifies the action of the inner product on the tensors of a single fixed locus.

The anomaly vectors $\vec{a},\vec{b}$ must lie on the charge lattice  \cite{Monnier:2017oqd}, and should be invariant under permutations. Thus, they should both be of the form
\begin{equation} \vec{a}=(\vec{v}_a,\vec{m}_a,\vec{m}_a,\ldots),\quad \vec{b}=(\vec{v}_b,\vec{m}_b,\vec{m}_b,\ldots),\end{equation}
which implies that the matrix of inner products
\begin{equation} \left(\begin{array}{cc}\vec{a}\cdot\vec{a}&\vec{a}\cdot\vec{b}\\\vec{a}\cdot\vec{b}&\vec{b}\cdot\vec{b}\end{array}\right)\,\equiv\,  \left(\begin{array}{cc}\vec{v}_a\cdot\vec{v}_a&\vec{v}_a\cdot\vec{v}_b\\\vec{v}_a\cdot\vec{v}_b&\vec{v}_b\cdot\vec{v}_b\end{array}\right)\, \text{mod}\, 8.\end{equation}
That is, the matrix of anomaly coefficients must be congruent modulo eight to the matrix of inner products of an even (not necessarily self-dual) lattice of signature $(1,1)$. This is a nontrivial constraint (similar to the one in \cite{Seiberg:2011dr}) that we can check case by case in all the examples that yield a non-abelian gauge group in six dimensions. We checked all the exceptional cases \footnote{With no quartic Casimir.} in table \ref{latabla} and found one inconsistent model, that of an $F_4$ gauge theory, since it has
\begin{equation} \left(\begin{array}{cc}\vec{a}\cdot\vec{a}&\vec{a}\cdot\vec{b}\\\vec{a}\cdot\vec{b}&\vec{b}\cdot\vec{b}\end{array}\right)=\left(\begin{array}{cc}0&3\\3&3\end{array}\right),\end{equation}
which is manifestly not embeddable in an even, self dual lattice. We already knew the $F_4$ theory is inconsistent \cite{Garcia-Etxebarria:2017crf}; it is nice that the anomaly analysis in I-fold compactifications precisely reproduces this fact. All the other examples with global anomalies in 6d lead to consistent inner product matrices.

\bibliographystyle{JHEP}
\bibliography{Ifolds-refs}

\end{document}